\documentclass{aastex62}
\usepackage{rotating}
\usepackage{xcolor}

\received{September 7, 2018}
\revised{October 8, 2018}

\submitjournal{PASP}

\shorttitle{Meteorological data from KLAWS-2G}
\shortauthors{Hu et al.}

\begin{document}

\title{Meteorological data from KLAWS-2G for an astronomical site survey of Dome A, Antarctica}

\def\mps{\,m\,s$^{-1}$}
\def\metre{\,m }
\def\arcseconds{\arcsec\ }
\def\cdeg{$^{\circ}$C}
\def\cgrad{$^{\circ}$C $\mbox{m}^{-1}$ }
\def\cgradd{$^{\circ}$C $\mbox{m}^{-1}$}

\correspondingauthor{Zhaohui Shang}
\email{zshang@gmail.com}

\author{Yi Hu}
\affil{National Astronomical Observatories, Chinese Academy of Sciences, Beijing, 100101, China}

\author{Keliang Hu}
\affil{National Astronomical Observatories, Chinese Academy of Sciences, Beijing, 100101, China}

\author{Zhaohui Shang}
\affil{Tianjin Normal University, Tianjin, 300387, China}
\affil{National Astronomical Observatories, Chinese Academy of Sciences, Beijing, 100101, China}

\author{Michael C. B. Ashley}
\affil{School of Physics, University of New South Wales, NSW 2052, Australia}

\author{Bin Ma}
\affil{National Astronomical Observatories, Chinese Academy of Sciences, Beijing, 100101, China}

\author{Fujia Du}
\affil{National Astronomical Observatories, Nanjing Institute of Astronomical Optics \& Technology,  Chinese Academy of Sciences, Nanjing, 210042, China}

\author{Zhengyang Li}
\affil{National Astronomical Observatories, Nanjing Institute of Astronomical Optics \& Technology,  Chinese Academy of Sciences, Nanjing, 210042, China}

\author{Qiang Liu}
\affil{National Astronomical Observatories, Chinese Academy of Sciences, Beijing, 100101, China}

\author{Wei Wang}
\affil{National Astronomical Observatories, Chinese Academy of Sciences, Beijing, 100101, China}

\author{Shihai Yang}
\affil{National Astronomical Observatories, Nanjing Institute of Astronomical Optics \& Technology,  Chinese Academy of Sciences, Nanjing, 210042, China}

\author{Ce Yu}
\affil{Tianjin University, Tianjin, 300072, China}

\author{Zhen Zeng}
\affil{National Astronomical Observatories, Chinese Academy of Sciences, Beijing, 100101, China}

\begin{abstract}

We present an analysis of meteorological data from the second generation of the
Kunlun Automated Weather Station (KLAWS-2G) at Dome A, Antarctica during 
2015 and 2016. We find that a strong temperature inversion exists 
for all the elevations up to 14\metre that KLAWS-2G can reach, and lasts for more than 
10 hours for 50\% or more of the time when temperature inversion occurs. The average 
wind speeds at 4\metre elevation are 4.2\mps\ and 3.8\mps\ during 2015 and 
2016, respectively. The strong temperature inversion and moderate wind speed 
lead to a shallow turbulent boundary layer height at Dome A. By analyzing the temperature 
and wind shear profiles, we note telescopes should be elevated by at least 8\,m above the ice. 
We also find that the duration 
of temperature inversions, and the wind speed, vary considerably from year to year. 
Therefore, long-term and continuous data are still needed for the site 
survey at Dome A. 

\end{abstract}


\section{Introduction} \label{sec:intro}

The Antarctic plateau has long been thought to contain the best sites on earth for many astronomical observations.
The most promising and well studies candidates include Dome~A, Dome~C, Dome Fuji, and Ridge~A.
By reviewing the available data for these sites from ground-based instruments, 
satellites and numerical simulations, \cite{Saunders09} and \cite{Burton10}
summarized their weather 
conditions, boundary layer heights, atmospheric seeing, clouds, aurorae, 
etc. These studies showed  excellent astronomical observing conditions at all four
Antarctic sites, superior in many cases to the 
best lower latitude sites, such as Mauna Kea. 

Long-term and in situ measurement data are essential to draw robust 
conclusions on the astronomical sites. For Dome~C, \cite{Lawrence04} 
reported a median free-atmospheric seeing of 0.23\arcseconds above a 
boundary layer just less than 30\metre high based on data from 
a multi-aperture scintillation sensor (MASS) and a sonic radar (SODAR). \cite{Aristidi05} 
analyzed two decades of temperature and wind speed data from an automated 
weather station and four summers of measurements with balloon-borne sondes, 
and found that Dome~C had an extremely stable upper atmosphere and a very 
low inversion layer. \cite{Travouillon08} used balloon-borne measurements 
to find that the median value of the boundary layer height at Dome C is 
33~m. \cite{Aristidi09} obtained a median seeing of 1.67\arcseconds at 3\,m, 
dropping to 0.84\arcseconds at 20\metre based on 3.5 years of differential image motion monitor (DIMM) 
data.

Up until now, only Dome~C among the sites mentioned above has had a regular manned operation over winter. 
Because of logistical difficulties, the other potential 
sites lack long-term and continuous data.

For Dome A, \cite{Bonner10} reported a median boundary layer height of 
13.9\metre using seven months of data obtained in 2009 from the Snodar instrument---a sonic
radar giving 1\,m resolution up to 200\,m above the ice.
By analyzing eight months of data 
from the first Kunlun automated weather station (KLAWS), \cite{Hu14} (hereafter HU14) 
found a strong and long lasting temperature inversion existing just above 
the snow surface, and an anti-correlation between the temperature 
inversion strength and the boundary layer height.

For Dome~Fuji, \cite{Okita13} reported a 0.53\arcseconds seeing at 11\metre, 
and a 0.2\arcseconds seeing  when the boundary layer height was  lower than 
11\metre. However, the time span of their data was only 18 days during 
January 2013.

Mast-based automated weather stations (AWS), such as KLAWS, with multiple sensors 
at different elevations are important for site evaluation. Not only can a mast AWS
give direct information on the atmospheric turbulence and wind speed below 
the boundary layer (HU14), it can also provide crucial data for site 
simulations \citep{Falvey16}. Moreover, it has helped us to 
operate unattended telescopes \citep{Shang12,Liu18} at Dome~A and the 
data are also useful for designing and building large diameter telescopes 
with adaptive optics systems \citep{Aristidi05} in the future.

In this paper, we present the results from analyzing continuous data spanning one 
year and eight months from KLAWS-2G. Instrumentation and sensor calibration 
are described in Section \ref{sec:ins&calib}. A statistical analysis of 
temperature, wind speed, and relative humidity data and a comparison with data from the first generation AWS are 
presented in Sections \ref{sec:results} and \ref{sec:compare}.  Finally, a 
discussion and summary is presented in Section \ref{sec:summary}.

\section{Instrument and calibration} \label{sec:ins&calib}

\subsection{Instrumentation}\label{subsec:ins}

\begin{figure}
\epsscale{0.5}
\plotone{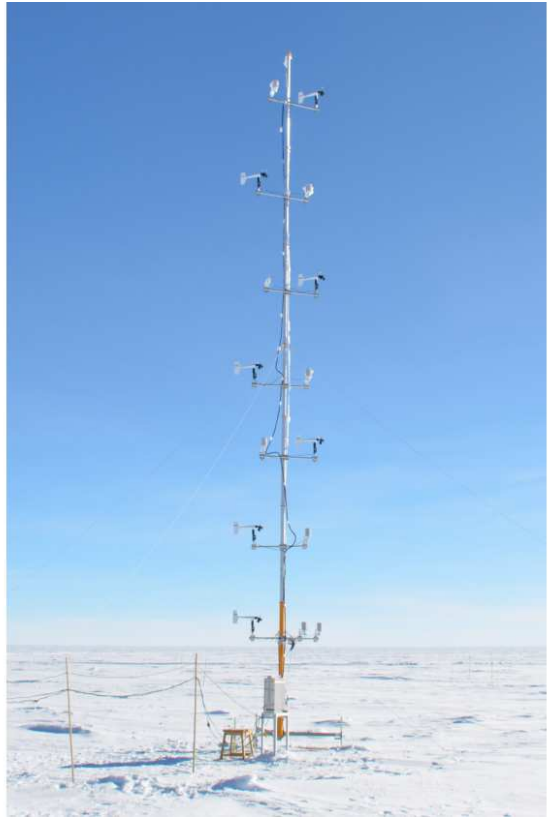}
\caption{KLAWS-2G installed at Dome~A in January 2015. The temperature 
sensors (in the radiation shields) and anemometers are clearly visible. 
The electronic box sits at the foot of the mast. The photo was taken in 
2016. \label{fig:klaws2g}}
\end{figure}

Our second generation weather station KLAWS-2G (see Figure \ref{fig:klaws2g}) was installed at Dome~A in 
January 2015 by the 31st CHINARE (Chinese National Antarctica Research 
Expedition) team.
KLAWS-2G has a 15\metre high mast, the same height as that of the first generation KLAWS (HU14). The mast supports ten 
temperature sensors, at heights of $-1$\,m (i.e., below the ice level), 0, 1, 2, 4, 6, 8, 10, 12 and 14\,m, seven wind 
speed and direction sensors (at 2, 4, 6, 8, 10, 12 and 14\,m), one air pressure 
sensor (at 2\,m), and one relative humidity sensor (at 2\,m). Temperatures are 
measured with 4-wire resistance temperature detectors (RTD) (Young Model 
41342). Wind speeds and directions are measured with propeller anemometers 
(Young Wind Monitor-AQ model 05305V). Air pressure is measured with a 
barometric pressure sensor (Young Model 61302V). Relative humidity is 
measured with a relative-humidity/temperature probe (Young Model 41382 VC). 
A custom data acquisition electronic box sits at the foot 
of the mast. The electronic box is connected to the controlling and 
operating system of the Antarctic Survey Telescope (AST3) via an RS232 cable 
\citep{Shang12}. The main computer of AST3 acquired the data from all the sensors every 10 
seconds and transferred them to the data server at our institute every 15 
minutes via Iridium satellite. The temperatures are read directly from the sensors, and other 
measurements are read as voltage values and then transformed to their 
actual values using the formulae given by the manufacturer. Power and internet connectivity for 
KLAWS-2G, and a warm operating environment for the computers, was provided by Plateau Observatory A (PLATO-A), an 
automated observatory \citep{Lawrence08} installed in 2012.

KLAWS-2G operated continuously from January 2015 to August 2016. 
During that time, all the sensors worked well apart for the barometer 
which suffered from ice accumulation. Some other problems developed over 
time. The radiation shield of the temperature sensor at 10\metre was broken 
since October 3, 2015 for there was an abrupt change in the behavior of the 
sensor. Later at the end of August 2016, we found that the temperature 
inversion disappeared above 6\metre and our webcam of KLAWS-2G saw the 
mast was bent at about 5\,m, indicating damage to the mast and some sensors. 
KLAWS-2G stopped working completely on September 9, 2016 caused by the 
failure of the serial port server device. In January 2017, KLAWS-2G was 
maintained by the 33rd CHINARE team, the serial port server device was 
replaced by a new one and all the sensors at 4\metre and below were 
resumed. The team also installed a new identical electronic box of KLAWS-2G 
to replace the old one. At this time, we placed a new barometer (Vaisala 
PTB2010 B2B4A) inside the electronics box rather than directly exposing it 
outside. From January 2017 until PLATO-A's kerosene supply ran out in May 2018, 
all the sensors up to 4\metre, including the new barometer, worked very well.

\subsection{Calibration}\label{subsec:calib}

Platinum RTDs are widely used in weather stations due to their stability and 
accuracy. Converting from resistance to temperature for a ``standard'' RTD 
is usually done using the Callendar-Van Dusen equation, or, for the highest accuracy, using the 
international standard ITS-90 \citep{Preston-Thomas90}. The measurement 
range defined in the standard reaches from $-259$\cdeg\ to $+961^{\circ}$C.
Our Young 41342 temperature sensor slightly differs from the 
standard RTD, and the manufacturer specifies a measurement range from
$-50$\cdeg\ to $+50$\cdeg. Since the 
actual temperature at Dome A was usually considerably below $-50$\cdeg, especially in 
wintertime (see HU14), we
sent three Young 41342 sensors to National Institute of 
Metrology, China (NIM) for calibrating in 2016. One of the three 
temperature sensors was purchased in 2014 and  was from the same batch of 
sensors used for KLAWS-2G. The other two were purchased in 2015 
from another production run. These calibrations allowed us to investigate both the 
zero shift and consistency of Young 40342 Model temperature sensors when used outside their nominal working temperature range.

\begin{figure}
\epsscale{0.85}
\plotone{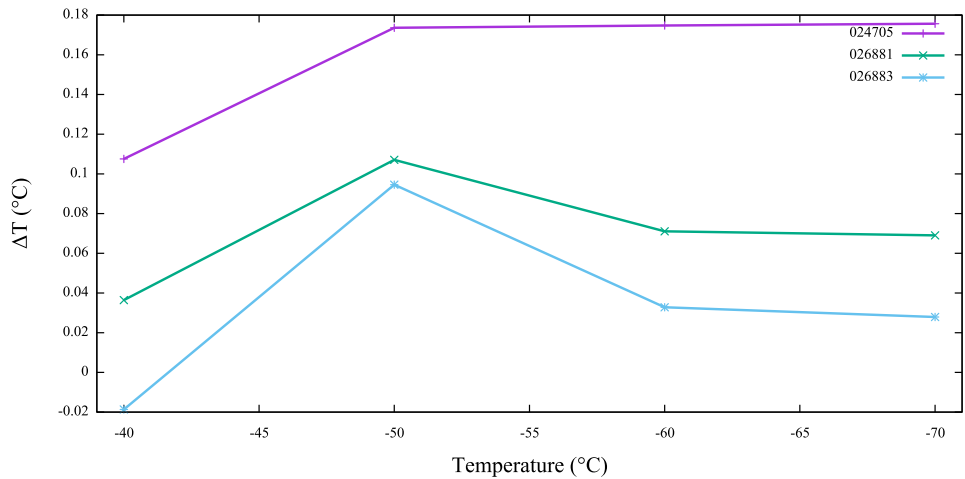}
\caption{Temperature differences between the actual temperature (column 3 
in Table \ref{tab:rtdcalib}) and the read temperature (column 5 Table 
\ref{tab:rtdcalib}).
\label{fig:rtdcalib}}
\end{figure}

The results from NIM include the requested calibrating temperatures, the 
actual calibrating temperatures and the measured resistances of our sensors, 
which can be converted to the temperatures our sensors would read. The 
calibration results are tabulated in Table \ref{tab:rtdcalib}. The actual 
calibration temperature (column 3) in each experiment was not strictly equal 
to that we requested (column 2). NIM used a standard high 
precision RTD to measure the actual temperature. All the expanded 
uncertainties of standard temperature are 0.02\cdeg, where the coverage factor 
$k=2$, which are also provided by NIM. Expanded uncertainty and coverage 
factor are explained in detail by \cite{Mohr16}. The temperatures measured 
by our temperature sensors can be calculated using the formula in Table 
\ref{tab:rtdcalib}. Figure \ref{fig:rtdcalib} shows the temperature 
differences between those from the standard RTD and those measured by these 
three sensors. The maximum difference is $+0.18$\cdeg\ under the 
$-70$\cdeg\ experiment, which is from the sensor we purchased in 2014. Comparing to the 
calibration results of the two sensors we purchased in 2015, the maximum 
difference is only 0.05\cdeg, while the maximum difference between the 
two batches is 0.15\cdeg. We note that the temperatures measured by the 
sensor purchased in 2014 are systematically larger than those measured by 
the sensors purchased in 2015, which might be caused by a zero shift. In 
conclusion, the consistency among the same batch of Young 41342 is better 
than 0.05\cdeg\ and the zero shift of this type of RTD between different 
batch should be less than 0.15\cdeg. The resistance to temperature 
formula can be safely expanded to $-80$\cdeg\ with a less than 
0.1\cdeg\ accuracy. Therefore we conclude that the Young Model 41342 is safe for using at Dome A. 

\begin{deluxetable*}{cCCCC}[htb!]
\tablecaption{Young Model 41342 temperature sensor calibration results 
\label{tab:rtdcalib}}
\tablecolumns{5}
\tablenum{1}
\tablewidth{0pt}
\tablehead{
\colhead{Serial No.} &
\colhead{Set} &
\colhead{Actual} &
\colhead{Resistance} &
\colhead{Read} \\
\colhead{} &
\colhead{(\cdeg)}  &
\colhead{(\cdeg)}  &
\colhead{($\Omega$)} &
\colhead{(\cdeg)}\\
\colhead{(1)} &
\colhead{(2)}  &
\colhead{(3)}  &
\colhead{(4)} &
\colhead{(5)}
}
\startdata
024705 & $-40.0$ & $-40.016$ & $846.921$ & $-39.91$ \\
(2014)   & $-50.0$ & $-50.008$ & $808.526$ & $-49.83$ \\
             & $-60.0$ & $-59.997$ & $769.761$ & $-59.82$ \\
             & $-70.0$ & $-70.008$ & $730.777$ & $-69.83$ \\
\hline
026881 & $-40.0$ & $-40.014$ & $846.654$ & $-39.98$ \\
 (2015)  & $-50.0$ & $-50.006$ & $808.276$ & $-49.90$ \\
             & $-60.0$ & $-59.997$ & $769.358$ & $-59.93$ \\
             & $-70.0$ & $-70.007$ & $730.365$ & $-69.94$ \\
\hline
026883 & $-40.0$ & $-40.019$ & $846.422$ & $-40.04$ \\
 (2015)  & $-50.0$ & $-50.010$ & $808.212$ &  $-49.92$ \\
             & $-60.0$ & $-59.995$ & $769.217$ &  $-59.96$ \\
             & $-70.0$ & $-70.012$ & $730.185$ & $-69.98$ \\
\enddata
\tablecomments{
Column 1 is the sensor serial number and date of manufacture.
Column 2 is the temperature that we requested for each experiment. 
Column 3 is the actual temperature of the experiment as measured by NIM's standard RTD. 
Column 4 is the resistance measured by NIM. 
Column 5 is the temperature converted 
from column 4 using the transfer function provided by the manufacturer: 
$T(^{\circ}C)=A{\cdot}R^2(\Omega)+B{\cdot}R(\Omega)+C$, 
$A=1.1279{\times}10^{-5}$, $B=2.3985{\times}10^{-1}$, $C=-251.1326$, 
where T is the temperature in Celsius, and R is the resistance in ohms.}
\end{deluxetable*}

\begin{figure}
\epsscale{0.8}
\plotone{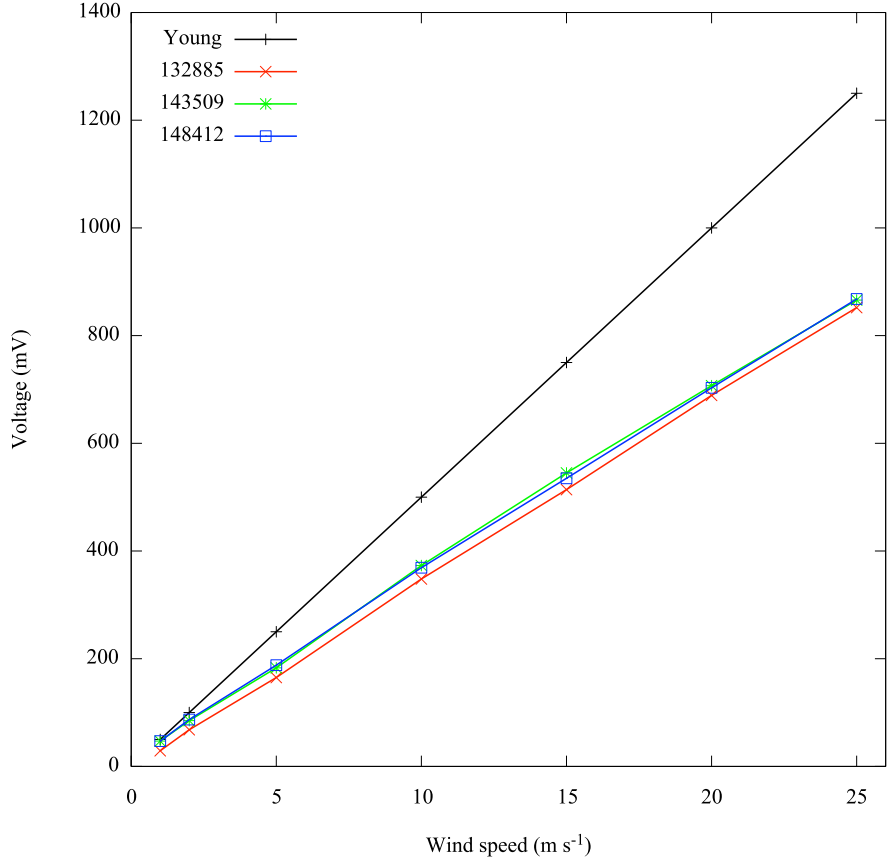}
\caption{Calibration results of the anemometers. The black line labeled as 
``Young'' is the transfer function in Table \ref{tab:anemcalib}. The data 
also come from Table \ref{tab:anemcalib}.\label{fig:anemcalib}}
\end{figure}

We also sent three anemometers to NIM for calibration. Since the size of the wind tunnel available at NIM was 
comparable to the anemometer size, we 
did not obtain absolute calibration results. However, we were still able to 
confirm the linearity of the transfer function of the Young 05305V anemometers. 
The calibration results are tabulated and showed in Table 
\ref{tab:anemcalib} and in Figure \ref{fig:anemcalib}. As showed in Figure 
\ref{fig:anemcalib}, except for the lowest two wind speeds, the voltages 
measured under the same experiments are all roughly two-thirds of 
the expected values calculated using the conversion formula given by 
the manufacturer.  This is caused by the effect of the air stream in the wind 
tube being partially blocked by the body of the anemometer. 
Throughout this text, we will use the transfer function provided by the manufacturer
to obtain the actual wind speed.

\begin{deluxetable*}{crrr}[htb!]
\tablecaption{Young Model 41342 wind speed sensor calibration results 
\label{tab:anemcalib}}
\tablecolumns{4}
\tablenum{2}
\tablewidth{0pt}
\tablehead{
\colhead{Serial No.}&
\colhead{Wind speed}&
\colhead{Measured}&
\colhead{Expected}\\
\colhead{} &
\colhead{[\mps]}&
\colhead{[mV]}&
\colhead{[mV]}\\
\colhead{(1)}&
\colhead{(2)}&
\colhead{(3)}&
\colhead{(4)} 
}
\startdata
132885 & $1.00$ & $29$ & $50$  \\
(2014) & $2.00$ & $68$ & $100$  \\
       & $5.00$ & $165$ & $250$  \\
       & $10.00$ & $348$ & $500$  \\
       & $15.00$ & $514$ & $750$  \\
       & $20.00$ & $689$ & $1000$  \\
       & $25.00$ & $852$ & $1250$  \\
\hline
143509 & $1.00$ & $47$ & $50$  \\
(2015) & $2.00$ & $85$ & $100$  \\
       & $5.00$ & $183$ & $250$  \\
       & $10.00$ & $373$ & $500$  \\
       & $15.00$ & $545$ & $750$  \\
       & $20.00$ & $707$ & $1000$  \\
       & $25.00$ & $866$ & $1250$  \\
\hline
148412 & $1.00$ & $47$ & $50$  \\
(2015) & $2.00$ & $87$ & $100$  \\
       & $5.00$ & $188$ & $250$  \\
       & $10.00$ & $396$ & $500$  \\
       & $15.00$ & $535$ & $750$  \\
       & $20.00$ & $703$ & $1000$  \\
       & $25.00$ & $868$ & $1250$  \\
\enddata
\tablecomments{Column 4 is the expected voltage converted from 
column 2 using the transfer function provided by the manufacturer: 
$V=kU$, $k=0.02$, where V is the wind speed in \mps, and U is the 
voltage in mV.}
\end{deluxetable*}

\subsection{Data caveats}\label{subsec:caveat}

We note a few problems in the data that readers should be aware of in 
understanding the data and results below.

The radiation shield of the temperature sensor at 10\metre dropped off on 
October 3, 2015. After this time, the temperature measured by the sensor at 
10\metre is higher than the true value, as we will show in Section 
\ref{sec:temp}.  We use a corrected temperature at 10\metre after 1 October 2015 
for calculating temperature gradient at heights of 10\metre and 12\metre (see Section \ref{sec:temp}).

According to the anemometer's user manual, 
there is a narrow blind spot in wind direction to the north. Also, because of blocking 
by the mast body, the measured wind speed, when the wind direction is near due
east or west, is smaller than the true speed, as we will show in Section 
\ref{sec:wind}. This is also why there is a gap in the wind rose plots. 

The relative humidity sensor's working temperature is only specified to work from 
$-50$\cdeg\ to $+50$\cdeg. Therefore, the relative 
humidity could be underestimated from its true value when the ambient temperature is 
lower than $-50$\cdeg. There are more details on this issue in Section  \ref{sec:humidity}.

The barometer only worked correctly for one month during 2016, as the result of problems 
mentioned in Section\ref{subsec:ins}. In this paper, we 
do not present air pressure data.

\section{Data statistics and results}\label{sec:results}

\subsection{Temperature and temperature inversion} \label{sec:temp}

\begin{figure}
\epsscale{0.9}
\plotone{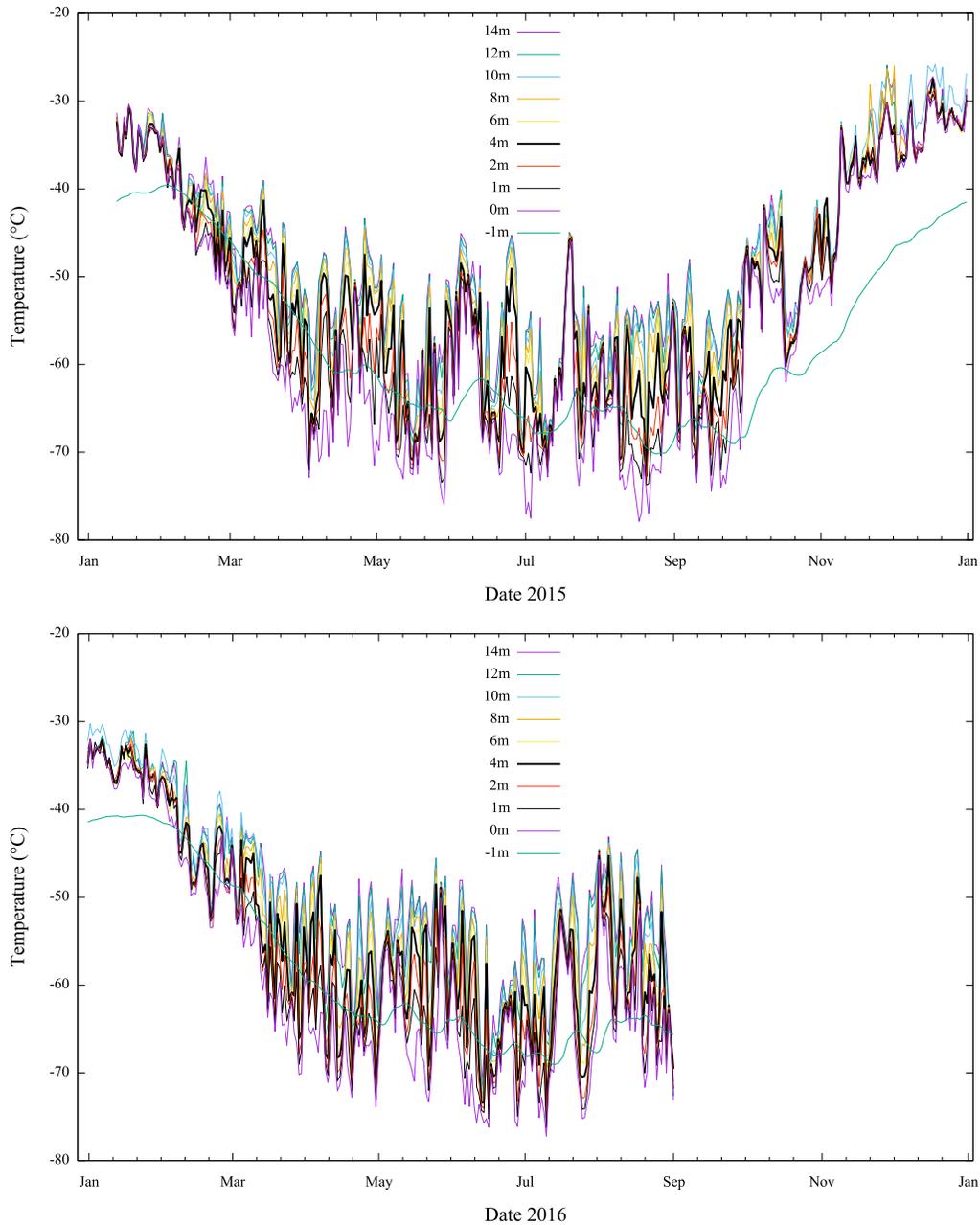}
\caption{Daily temperature median during 2015 and 2016. 
\label{fig:dailytemp}}
\end{figure}

To show the long-term trend within an entire year, we plot the daily median 
temperatures in 2015 and 2016 in Figure \ref{fig:dailytemp}. Taking 2\metre 
as an example, the daily median temperature at 2\metre in 2015 was 
approximately $-35$\cdeg\ from January to February, then dramatically went 
down to roughly $-60$\cdeg\ in April, and could frequently reach below 
$-70$\cdeg\ in the dark winter. It rose again after October when the polar 
night was over. The daily median temperature above the surface could vary 
more than 10\cdeg\ in two days, but the daily median temperature at 
$-1$\metre (1\metre below the surface) varied smoothly because thick loose 
snow covers the ground surface at Dome A and isolates heat transmission. 
Because of the good thermal insulation of the loose snow, the temperature 
at $-1$\metre is lower than the surface temperature in summer and higher in 
the winter (Figure~\ref{fig:dailytemp}). The same trend can also be seen 
in the 2011 data (see Figure 7 in HU14).

\begin{figure}
\plotone{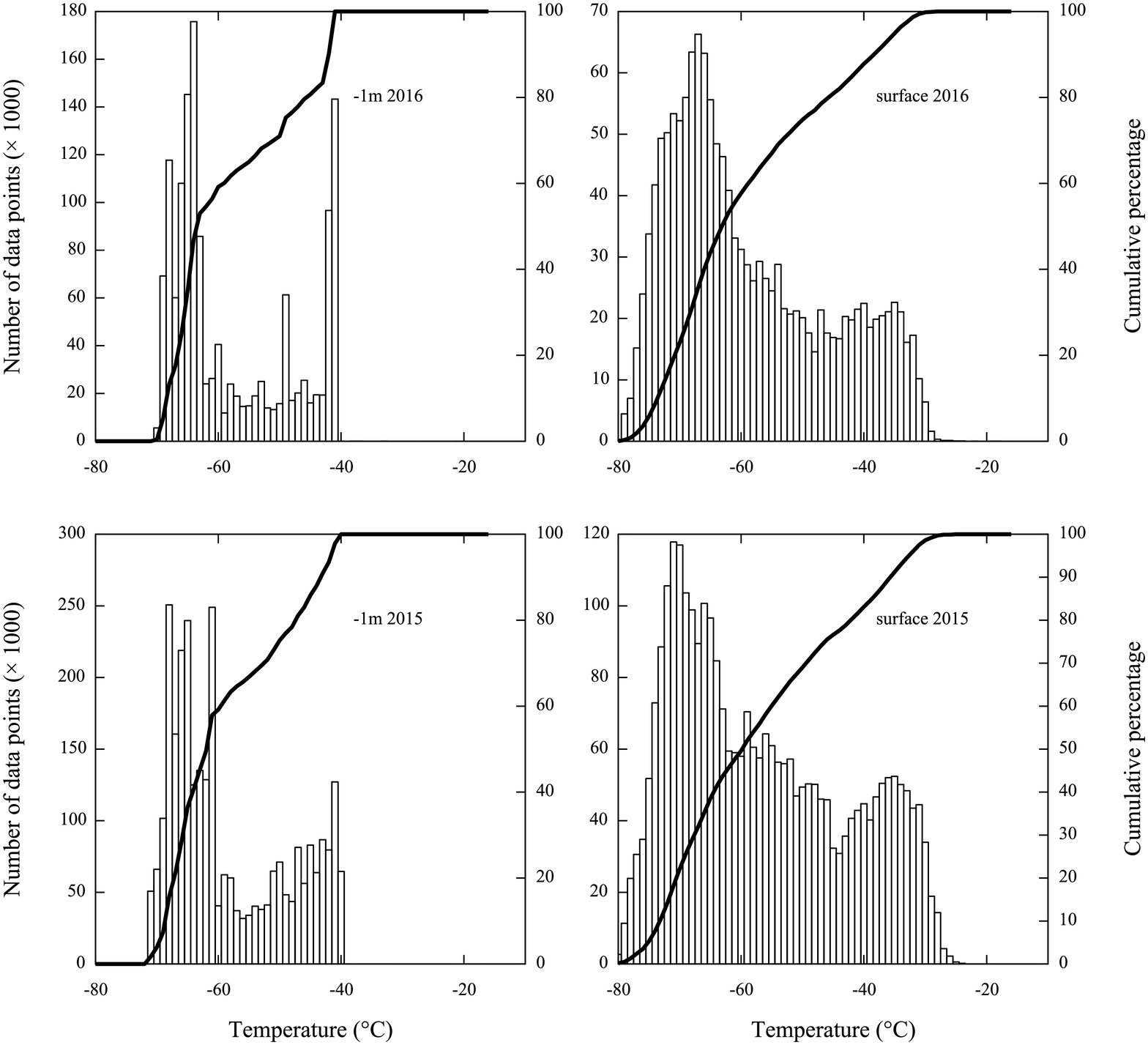}
\caption{Histograms and cumulative distributions (solid line) of the 
temperatures of surface and 1\metre under the surface during 2015 and 2016. 
\label{fig:tempdist1}}
\end{figure}

We show the detailed temperature distributions of $-1$\metre and surface 
(0\,m) in Figure \ref{fig:tempdist1}. The temperature at $-1$\metre in both 
years has a much smaller range (from $-70$\cdeg\ to $-40$\cdeg) than the air 
temperature (from $-80$\cdeg\ to -$25$\cdeg), as we had seen in 2011 (HU14). 
We note that the surface air temperature could reach below $-80$\cdeg\ in 
2015 and 2016. 

\begin{sidewaysfigure}
\epsscale{1.1}
\plotone{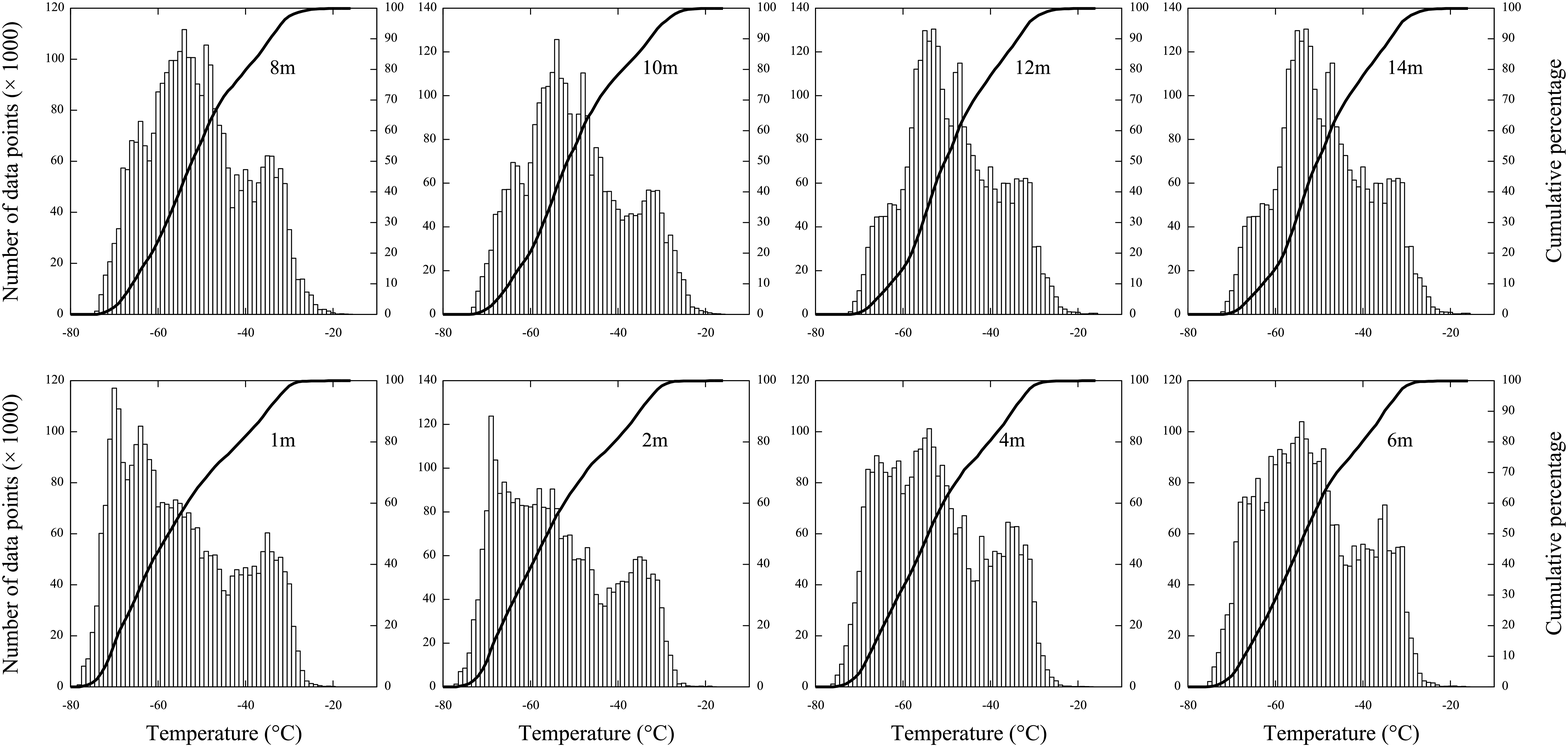}
\caption{Histograms and cumulative distributions (solid line) of the 
temperatures from  1\metre to 14\metre during 2015. \label{fig:tempdist2}}
\end{sidewaysfigure}

\begin{sidewaysfigure}
\epsscale{1.1}
\plotone{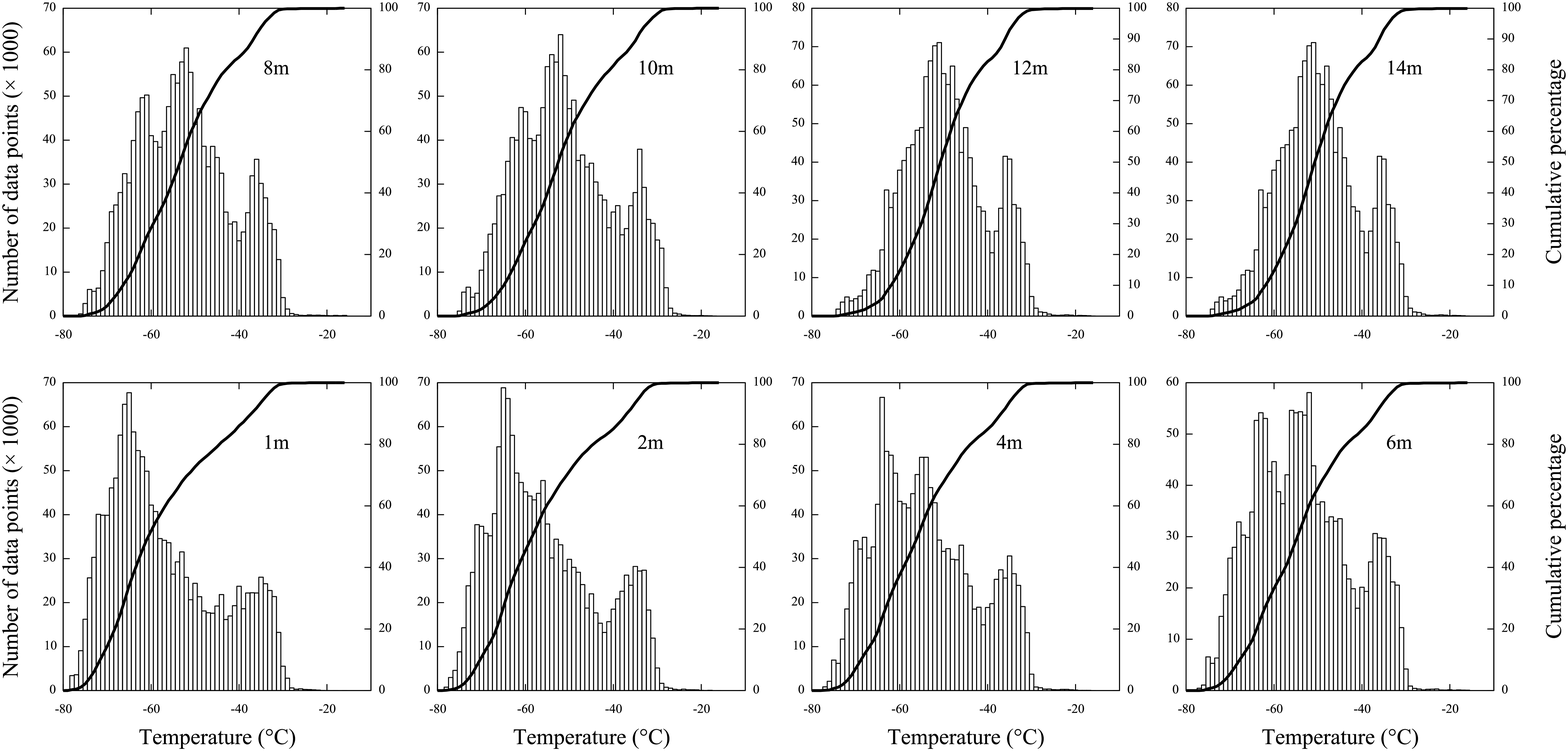}
\caption{Histograms and cumulative distributions (solid line) of the 
temperatures from 1\metre to 14\metre during 2016.  \label{fig:tempdist3}}
\end{sidewaysfigure}

The temperature distributions of the other elevations are shown in Figures 
\ref{fig:tempdist2} and \ref{fig:tempdist3}. We can see there are shallow 
valleys between $-50$\cdeg\ and $-40$\cdeg\ in these temperature distribution 
figures. It divides the data into three components: the long cold winter, 
warm summer, and the valley where the temperatures changed relatively 
rapidly during spring and autumn as we have seen in Figure 
\ref{fig:dailytemp}. The valleys are more obvious in 2016 than those in 
2015, because the data after September 2016 were absent, while the rapid 
changing of temperature in spring and autumn is not symmetric.

\begin{figure}
\epsscale{0.9}
\plotone{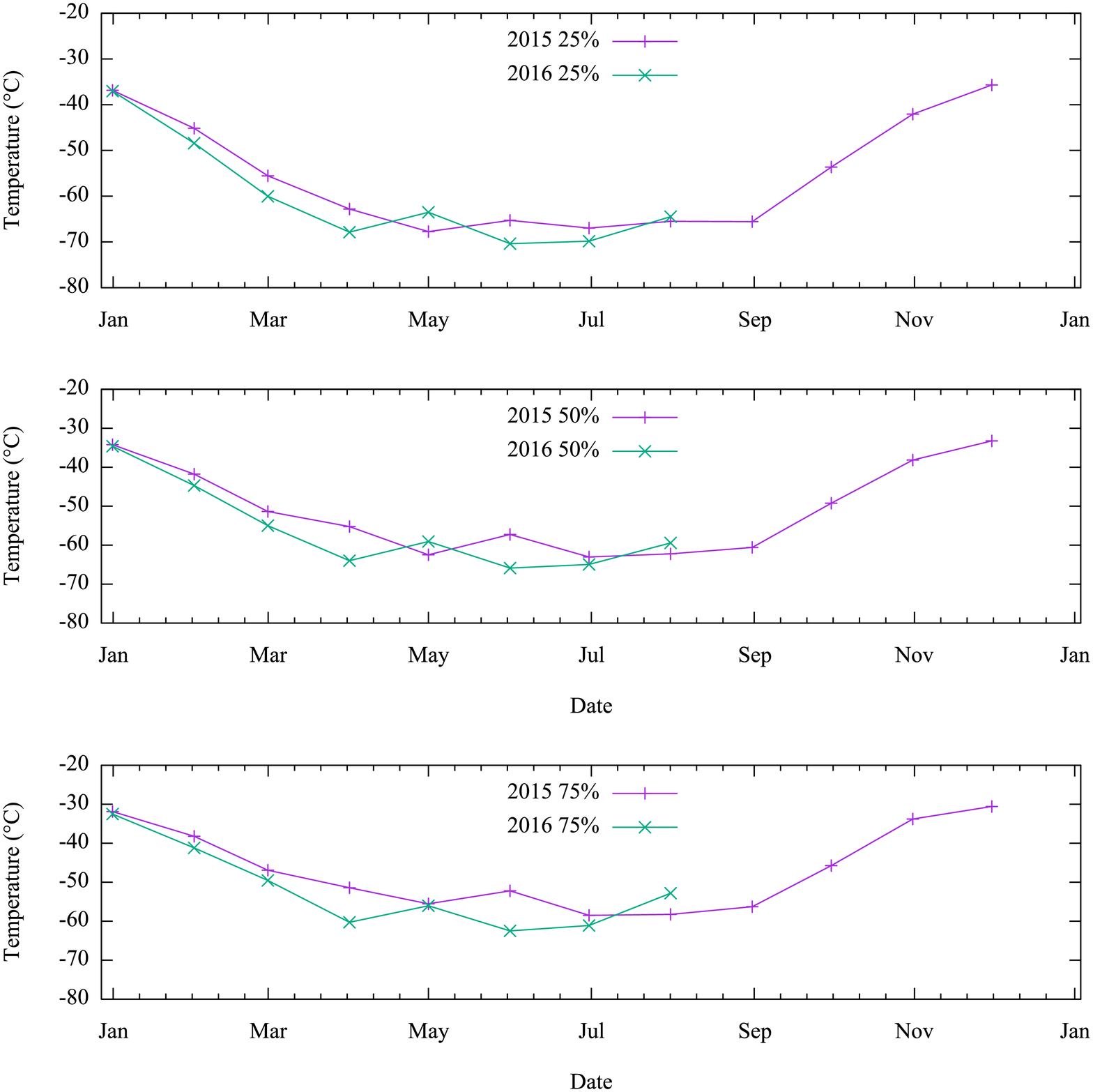}
\caption{Monthly 25\%, 50\%, and 75\% percentile temperature of 4\metre in 
2015 and 2016. \label{fig:monthlytemp}}
\end{figure}

To investigate annual variation, we list the 25\%, 50\% and 75\% monthly 
percentiles at 4\metre during 2015 and 2016 in Table \ref{tab:monthlytemp} 
and plot them in Figure \ref{fig:monthlytemp}. The monthly percentiles in 
2016 are lower than those in 2015, except in May and August. This suggests 
that the temperature in 2016 was lower than that in 2015, but the month-to-month
median temperature can vary as much as 5\cdeg.  

\setcounter{table}{2}
\begin{table}
\renewcommand{\thetable}{\arabic{table}}
\centering
\caption{Monthly percentile temperature at 4\metre during 2015 and 2016.} 
\label{tab:monthlytemp}
\begin{tabular}{cCCCCCCCC}
\tablewidth{0pt}
\hline
\hline
Month & \multicolumn2C{25\%} && \multicolumn2C{50\%} && \multicolumn2C{75\%}\\
\cline{2-3} \cline{5-6} \cline{8-9}
 &  (2015) &  (2016) &  &(2015) &  (2016) &&  (2015) &  (2016) \\
\hline
Jan & $-37.38$ & $-37.03$ && $-34.34$ & $-34.62$ && $-31.80$ & $-32.53$\\
Feb & $-46.80$ & $-48.42$ && $-43.23$ & $-44.71$ && $-39.26$ & $-41.20$\\
Mar & $-58.17$ & $-60.02$ && $-53.64$ & $-54.99$ && $-48.76$ & $-49.55$\\
Apr & $-64.38$ & $-67.86$ && $-57.98$ & $-63.98$ && $-54.35$ & $-60.29$\\
May & $-69.12$ & $-63.52$ && $-65.03$ & $-59.06$ && $-59.11$ & $-56.04$\\
Jun & $-66.50$ & $-70.38$ && $-60.64$ & $-65.88$ && $-54.83$ & $-62.49$\\
Jul & $-68.44$ & $-69.83$ && $-65.69$ & $-64.94$ && $-60.77$ & $-61.11$\\
Aug & $-68.42$ & $-64.48$ && $-64.98$ & $-59.44$ && $-61.14$ & $-52.84$\\
Sep & $-67.60$ & - && $-63.14$ & - && $-58.58$ & -\\
Oct & $-54.01$ & - && $-50.00$ & - && $-45.62$ & -\\
Nov & $-42.75$ & - && $-38.22$ & - && $-33.77$ & -\\
Dec & $-35.59$ & - && $-33.08$ & - && $-30.37$ & -\\
\hline
\end{tabular}
\end{table}

\begin{sidewaysfigure}
\epsscale{1.1}
\plotone{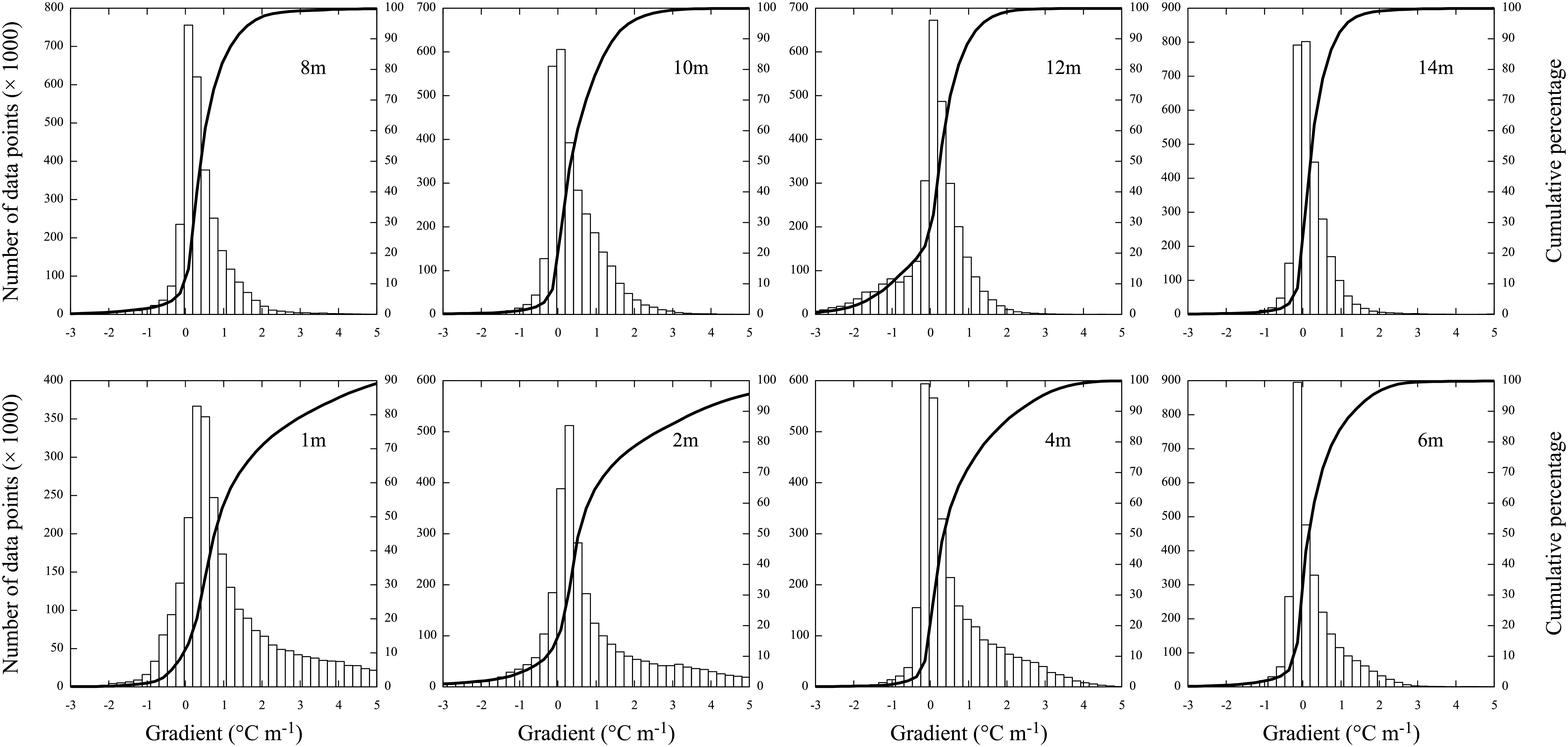}
\caption{Histograms and cumulative distributions (solid line) of the 
temperature gradients from 1\metre to 14\metre during 2015. The 
temperature gradients are calculated by corrected temperature at 10\metre
after 1 October 2015 (see Section \ref{sec:temp}).
\label{fig:tempgrad1}}
\end{sidewaysfigure}

\begin{sidewaysfigure}
\epsscale{1.1}
\plotone{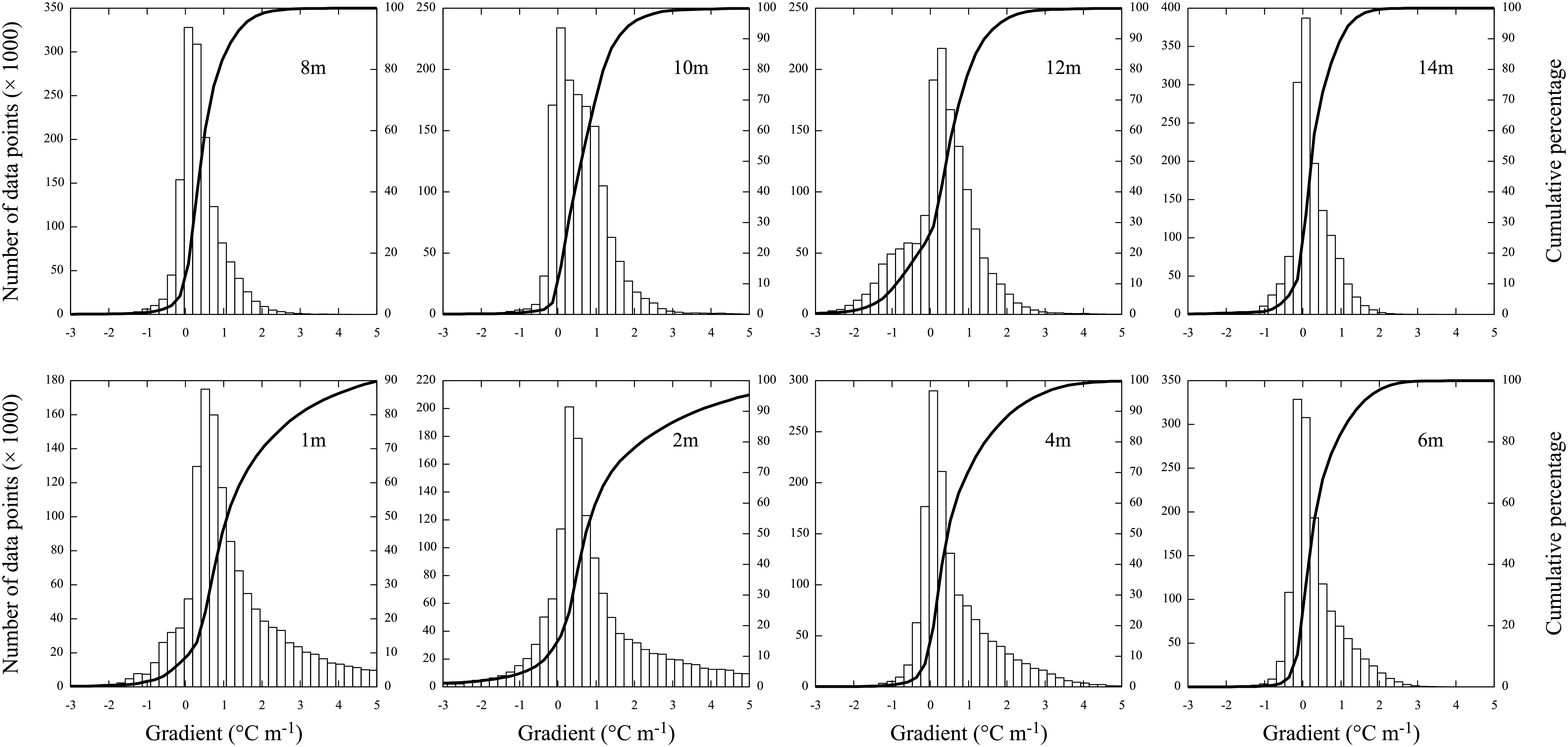}
\caption{Histograms and cumulative distributions (solid line) of the 
temperature gradients from 1\metre to 14\metre during 2016. 
The temperature gradients are calculated by corrected temperature at 10\metre.
 The temperature gradients are calculated by corrected temperature at 10\metre (see Section \ref{sec:temp}).
\label{fig:tempgrad2}}
\end{sidewaysfigure}

The temperature inversion (i.e., temperature decreasing with increasing elevation),
can be clearly seen in  Figure \ref{fig:dailytemp}, as we also saw in 2011 
(HU14). The temperature gradients are calculated and shown in Figures 
\ref{fig:tempgrad1} and \ref{fig:tempgrad2}. Note that we calculate the temperature gradient
at a given height from the temperature difference between that height and
the adjacent lower height, divided by the height difference. For example, the
gradient at 8\,m is $(T(8\,\mbox{m})-T(6\,\mbox{m}))/2$. A positive
gradient indicates a temperature inversion.

As we have  seen in Figure \ref{fig:dailytemp}, the temperature at 10\metre became the 
highest from November 2015 until March 2016 when the Sun started to set 
below the horizon. We noticed this in early October 2015 and inferred that 
this is not real and it could be that the radiation shield of the 
temperature sensor at 10\metre was broken or dropped off. This was 
confirmed by the 32nd CHINARE team in early 2016 but it could not be fixed 
on site. For this reason, hereafter we calculated the temperature gradient at 10\metre assuming 
that the variation was linear from 8 to 12 m. The plots all show a very strong temperature inversion 
that the gradient distributions increase rapidly from negative values to a 
narrow positive peak and go down with a relatively long tail.

\begin{figure}
\plotone{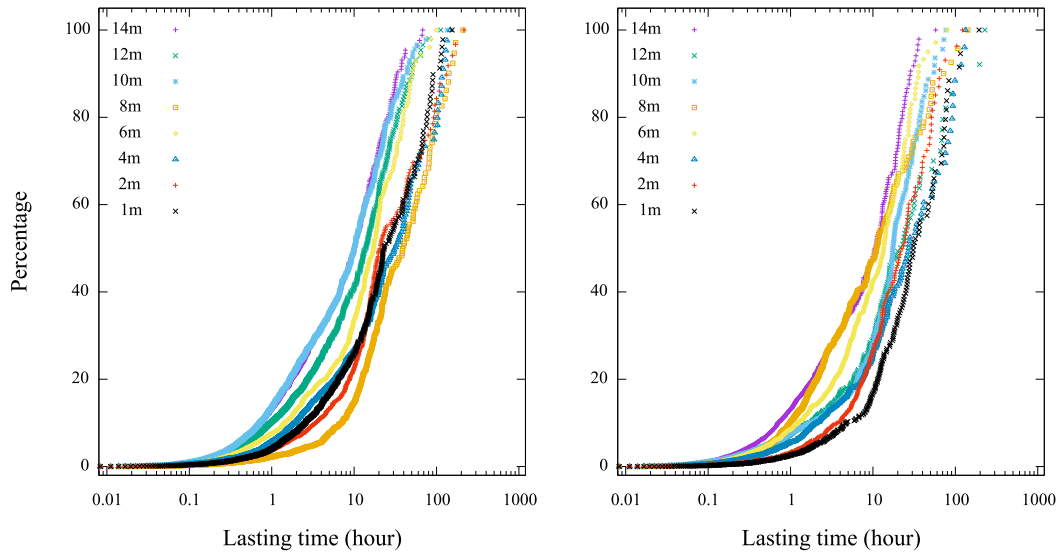}
\caption{Temperature inversion duration time during 2015 (left panel) and 
2016 (right panel). The 
temperature gradients are calculated by corrected temperature at 10\metre
after 1 October 2015. \label{fig:last}}
\end{figure}

When the temperature inversion happens, it can last a long time. To 
quantify how long the temperature inversion can last, we first define a 
real temperature inversion by adopting the criteria of temperature 
difference larger than 0.14\cdeg\ as  we did in Hu14. This is because that 
the temperature sensors are from the same batch of production, according to 
the calibration results presented in Section \ref{subsec:calib}, it is 
reasonable to regard that a temperature difference lager than 0.14\cdeg\ is 
a real measurement rather than random measurement error or systematic zero 
shift between two temperature sensors. Then the inversion lasting time is 
defined as a time period in which all the temperature differences are equal 
or larger than 0.14\cdeg. The cumulative distribution of temperature 
inversion lasting time are listed in Table \ref{tab:last} and shown in 
Figure \ref{fig:last}.  For all the heights, more than 50\% of the time, 
the temperature inversion could last more than 10 hours. The longest 
temperature inversion was 6 days at 4\metre elevation in 2015. 

\setcounter{table}{3}
\begin{table}
\renewcommand{\thetable}{\arabic{table}}
\centering
\caption{Cumulative percentages of the durations of temperature inversion.} \label{tab:last}
\begin{tabular}{cCCCCCCCC}
\tablewidth{0pt}
\hline
\hline
Height & \multicolumn2c{$>$10 hr} && \multicolumn2c{$>$25 hr} && \multicolumn2c{Total}\\
\cline{2-3} \cline{5-6} \cline{8-9}
(m) & \% (2015) & \% (2016) &  & \% (2015) & \% (2016) & &\% (2015) & \% (2016) \\
\hline
1  & 74.1 & 83.4 && 48.9 & 60.3 && 85.7 & 89.8 \\
2  & 77.5 & 73.7 && 45.3 & 47.1 && 79.2 & 81.8 \\
4  & 73.0 & 73.9 && 53.6 & 53.2 && 72.7 & 81.4 \\
6  & 66.4 & 57.8 && 34.9 & 25.1 && 56.6 & 66.8 \\
8  & 85.5 & 52.7 && 58.5 & 31.5 && 85.9 & 84.1 \\
10 & 45.0 & 79.8 && 18.0 & 58.8 && 63.1 & 88.0 \\
12 & 62.7 & 79.8 && 30.1 & 58.7 && 73.9 & 88.0 \\
14 & 50.0 & 51.9 && 20.1 & 15.2 && 66.5 & 69.3 \\
\hline
\end{tabular}
\end{table}

\subsection{Wind speed and wind direction}\label{sec:wind}

\begin{figure}
\plotone{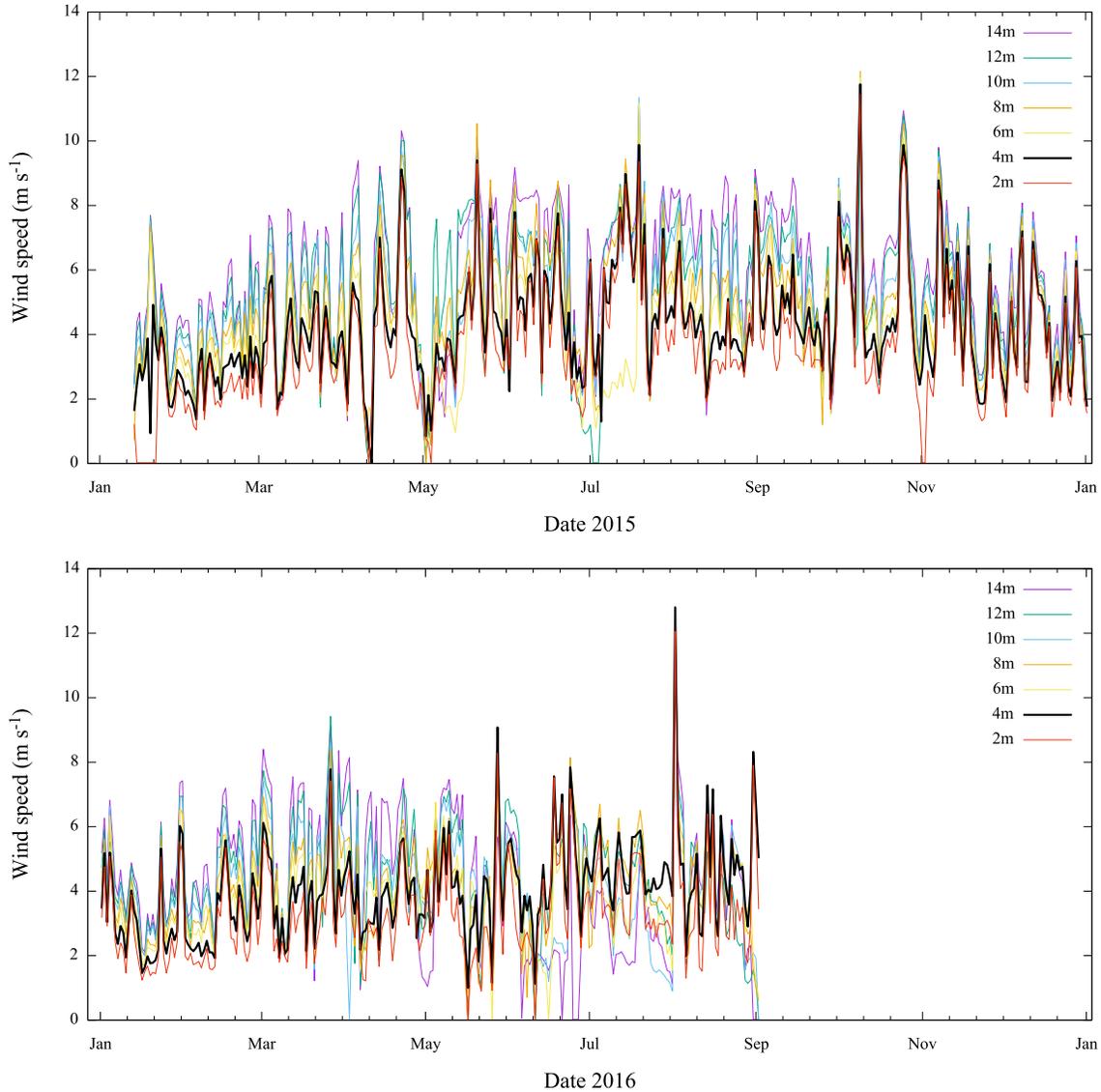}
\caption{Daily median wind speed during 2015 and 2016. \label{fig:dailyws}}
\end{figure}

The daily median wind speeds in 2015 and 2016 are shown in Figure 
\ref{fig:dailyws}. Unlike for the temperature, there were no obvious seasonal 
trends of wind speed. However, the wind speed in the winter season (from 
April to September) is larger than that in the summer season (from October to 
March). For example, the average wind speed at 4\metre in the summer of 
2015 was 4.0\mps, while it was 4.5\mps\ in the winter. The average wind 
speeds at 4\metre during 2015 and 2016 were 4.2\mps\ and 3.8\mps, 
respectively. In general, 
higher elevations have higher wind speeds. Sometimes, the daily wind speed 
was recorded as zero (see Figure \ref{fig:dailyws}). This occurred, for example, during a few days in July 2015 for the sensor at 
12\metre. We infer that such extended periods of zero readings are usually not real, but 
probably caused by the anemometer being frozen after a period of very low 
wind speed.  As seen in Figure \ref{fig:dailyws}, when the wind speed at a 
certain elevation is almost zero and those at other elevations were not 
zero at the same time, we always found that after a strong wind, the 
anemometer with problem could recover, indicating that it got stuck 
(possibly frozen) during a low wind speed period.

\begin{sidewaysfigure}
\epsscale{1.1}
\plotone{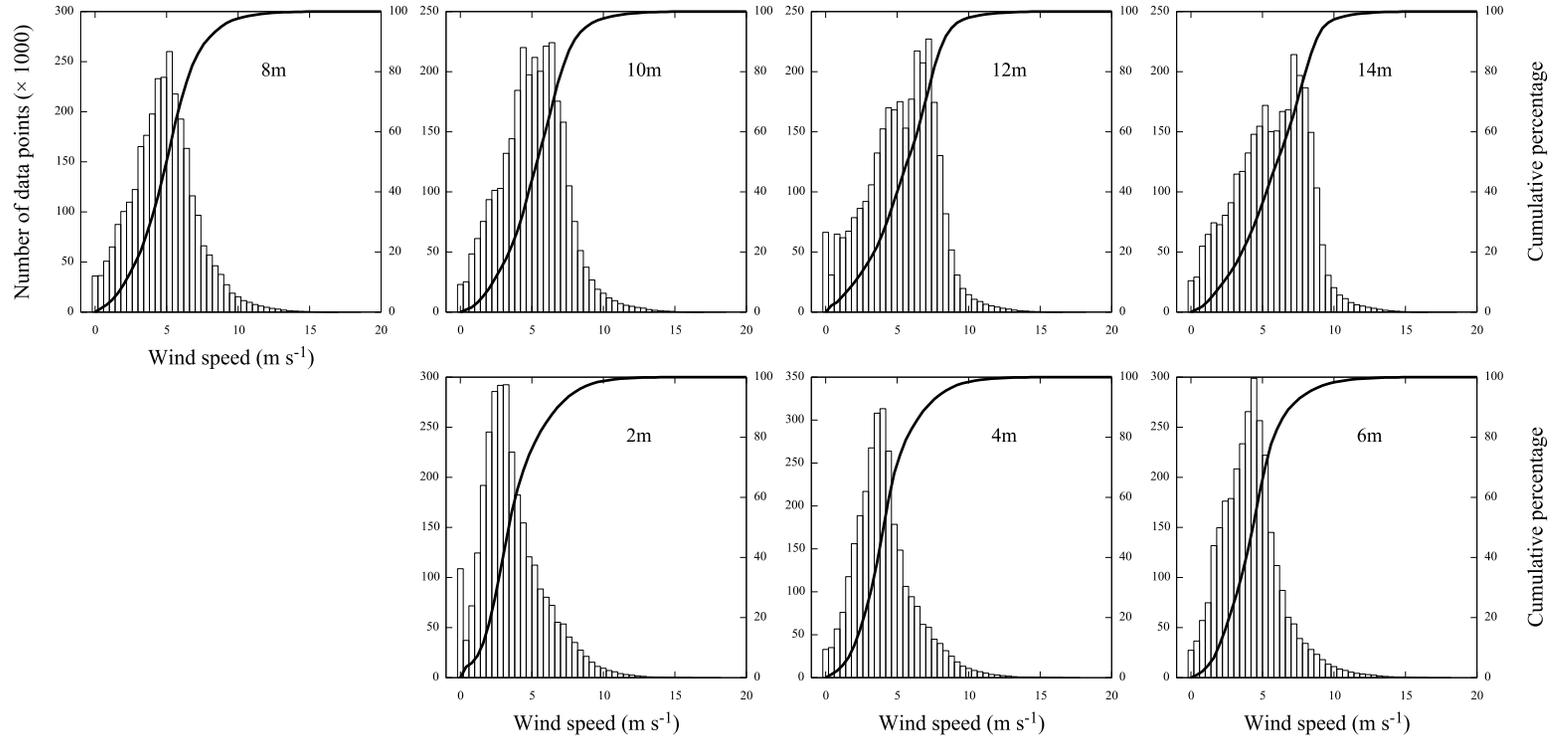}
\caption{Histograms and cumulative distributions (solid line) of wind 
speeds from 2\metre to 14\metre during 2015. \label{fig:wsdist2015}}
\end{sidewaysfigure}

\begin{sidewaysfigure}
\epsscale{1.1}
\plotone{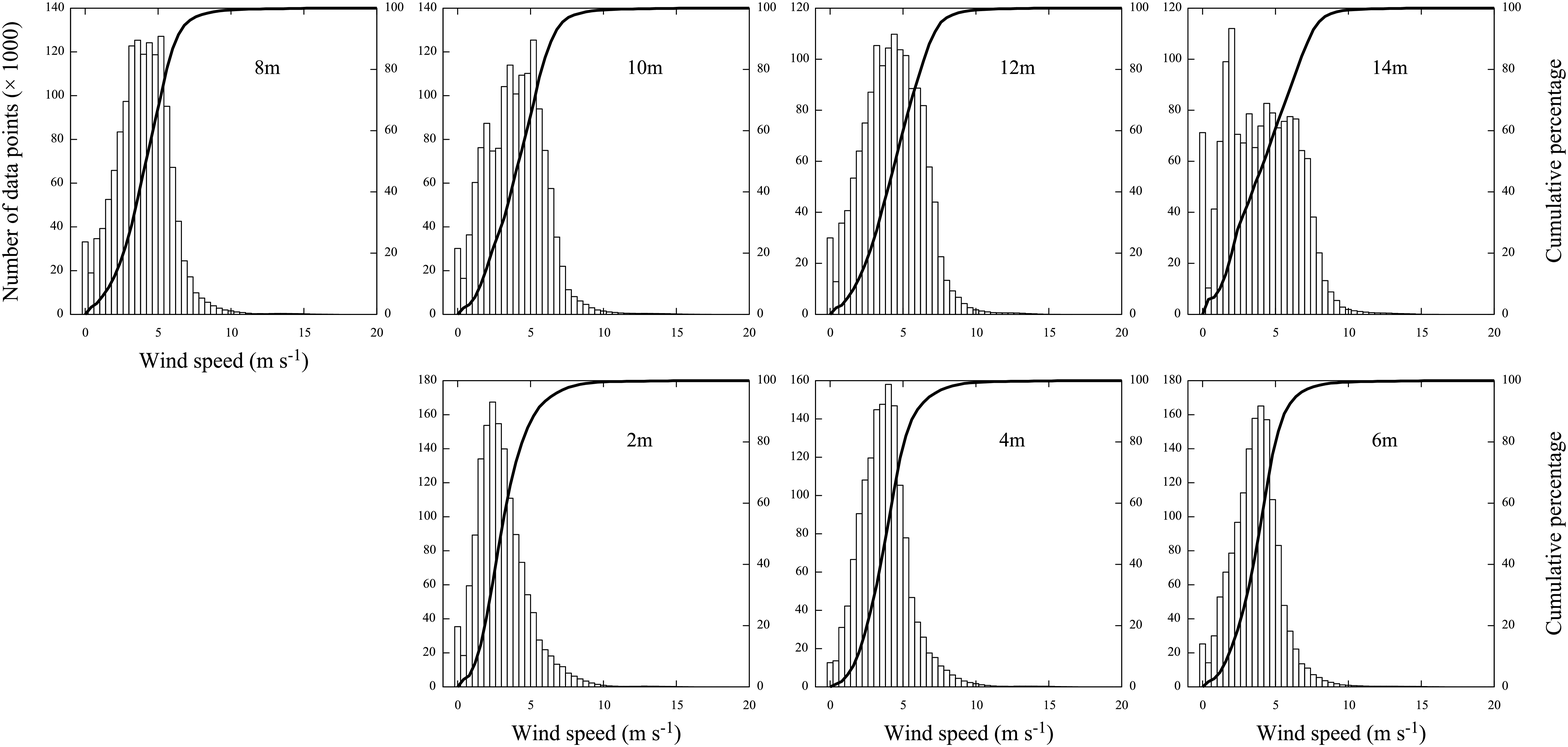}
\caption{Histograms and cumulative distributions (solid line) of wind 
speeds from  2\metre to 14\metre during 2016. \label{fig:wsdist2016}}
\end{sidewaysfigure}

The wind speed distribution during 2015 and 2016 are shown in Figures 
\ref{fig:wsdist2015} and \ref{fig:wsdist2016}. The distributions of wind 
speed at all the elevations show an asymmetric peak, whose position depends 
on its corresponding height. 

We note that there could be more low wind speed points, especially the zero 
wind speed points, than there should be, because the anemometer might be partially 
frozen when the wind speed is low, but we cannot quantify this effect. 

Although a higher elevation has a higher wind 
speed, the wind speeds for all the elevations seldom exceeded 10\mps, which 
is a great advantage for astronomical observatories. 

\begin{figure}
\epsscale{1.1}
\plotone{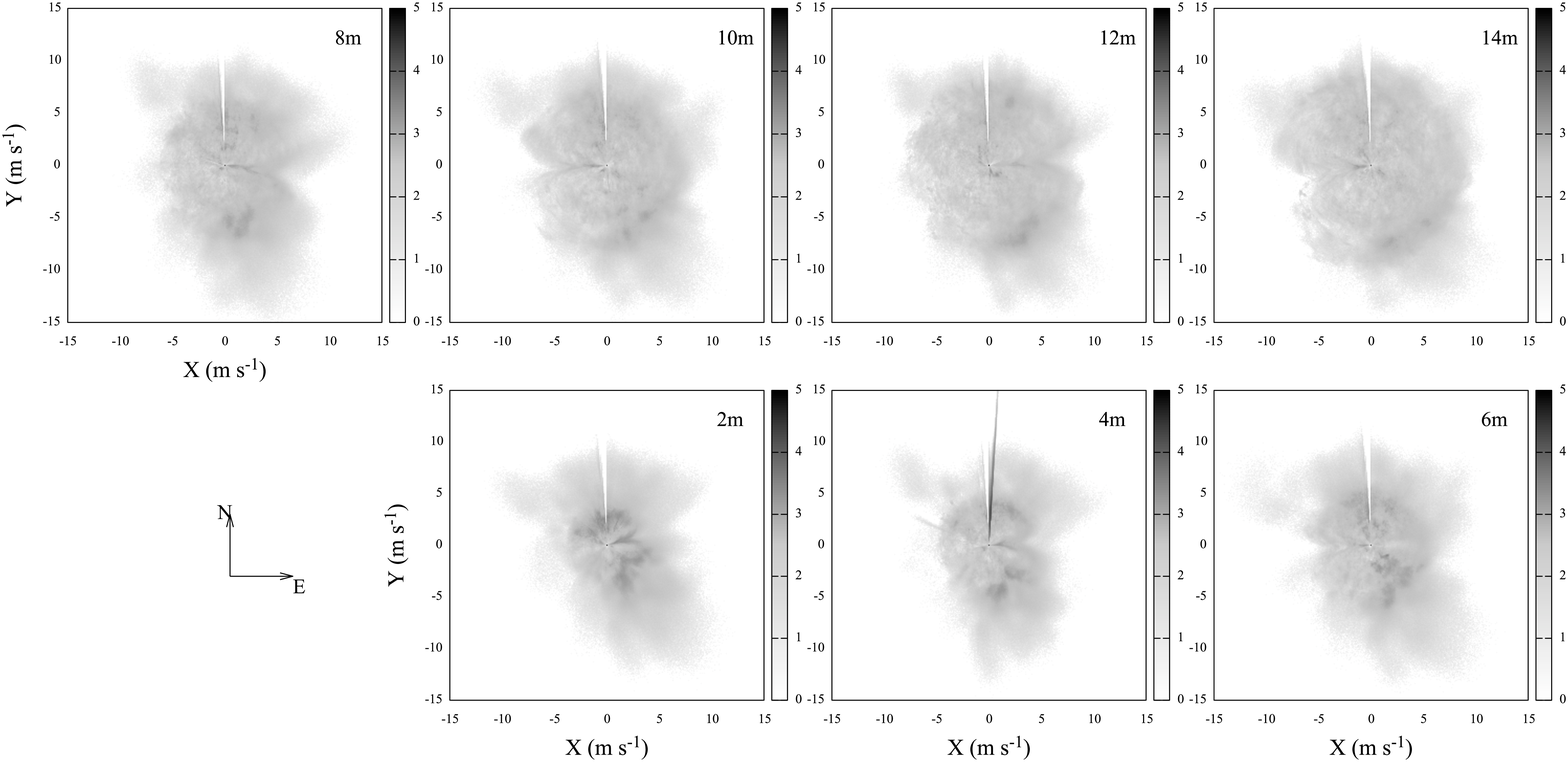}
\caption{Wind rose density at seven heights during 2015. The grey scale is 
the logarithm of the number of data points in each pixel. The blind strips 
to the north are caused by an imperfection in the acquisition module (see text in 
Section \ref{sec:wind} for details). The gaps to the east or west are caused 
by mast blocking.\label{fig:wr2015}}
\end{figure}

\begin{figure}
\epsscale{1.1}
\plotone{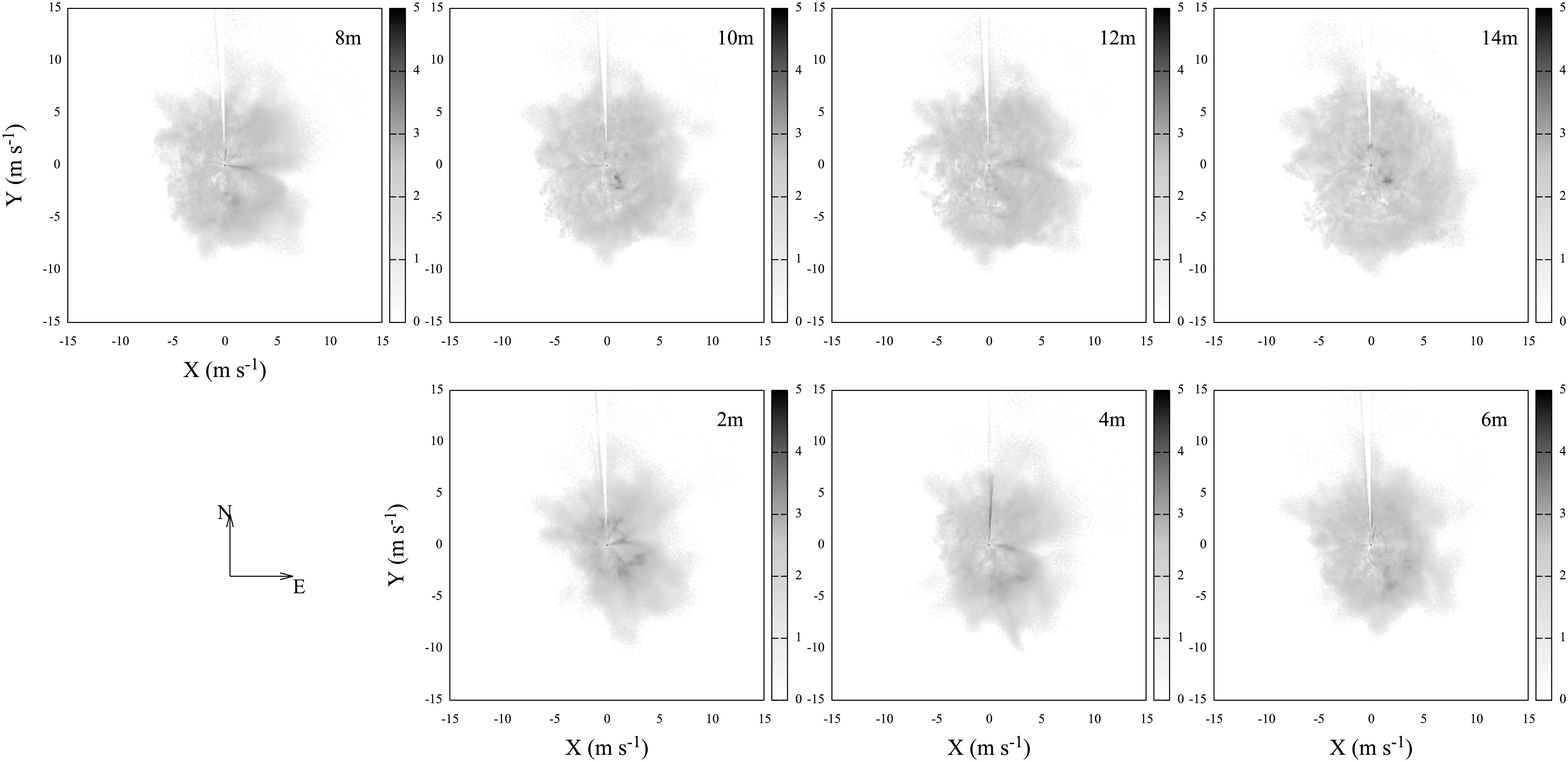}
\caption{The same figure as Figure \ref{fig:wr2015} during 2016. 
\label{fig:wr2016}}
\end{figure}

The wind rose plots in 2015 and 2016 are shown in Figure \ref{fig:wr2015} 
and \ref{fig:wr2016}. There is a very narrow ``blind'' spot to the north in wind 
direction as we mentioned in Section \ref{subsec:caveat}. There are also 
gaps to the east or west directions which are caused by mast blocking.  The direction 
of the gap is depended on position of the corresponding anemometer. The 
wind rose density plots show there is a slightly preferred southeast wind 
direction, at about 150\degr. 

 \subsection{Relative humidity}\label{sec:humidity}

According to the World Meteorological Organization, relative humidity is 
defined as the ratio of water vapor pressure and saturated water vapor 
pressure over water under that temperature, regardless of whether the 
temperature is larger than the freezing point of water \citep{Hardy98a}. Our 
relative humidity sensor gives measurements conforming to WMO's definition, 
according to its manual. However, we did not calibrate our humidity sensor 
due to the lack of a suitable experimental chamber that can provide a very 
dry, low air pressure and low-temperature environment. 

\begin{figure}
\plotone{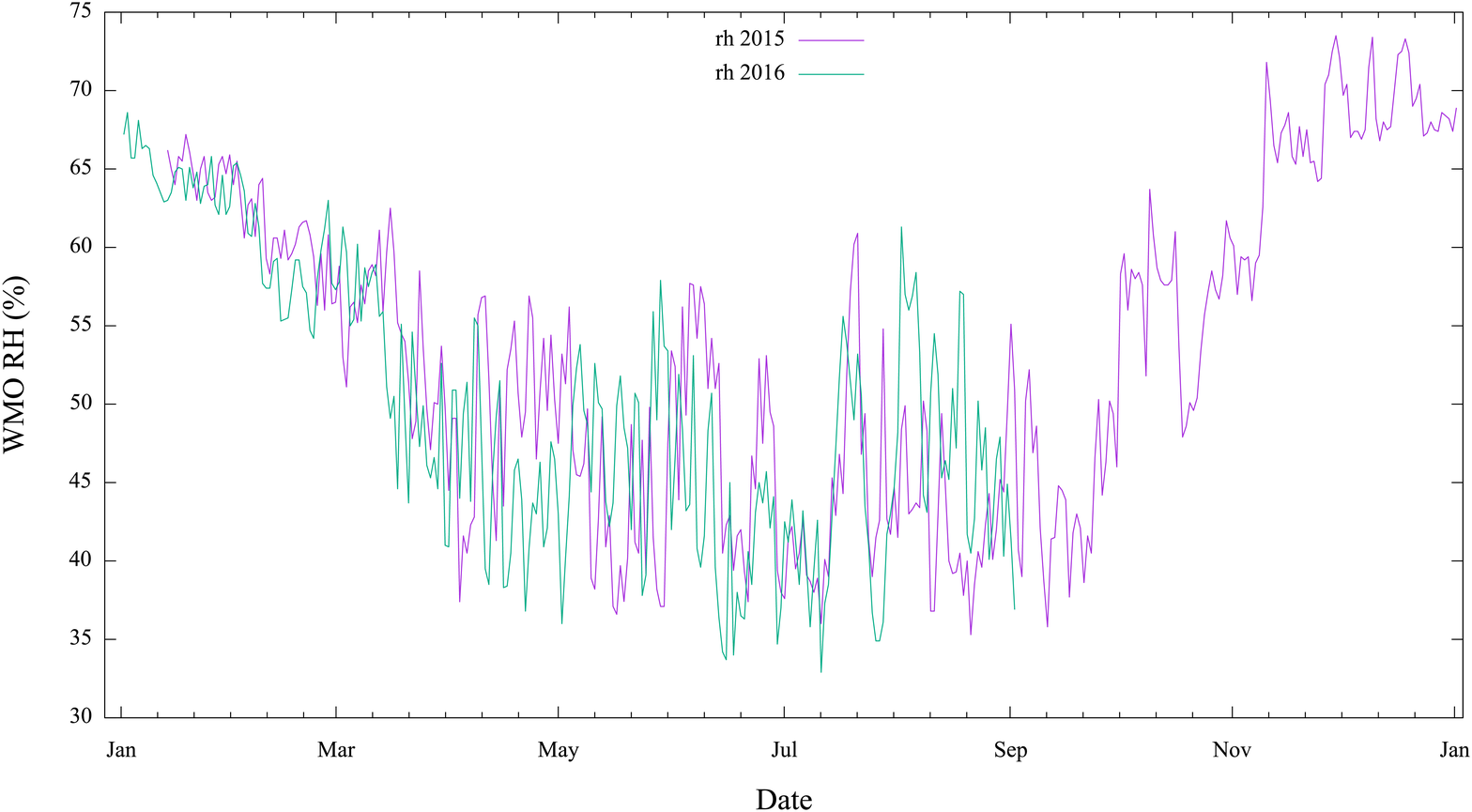}
\caption{Daily median WMO RH during 2015 and 2016. 
\label{fig:dailyrh}}
\end{figure}

Figure \ref{fig:dailyrh} shows the daily median relative humidity during 
2015 and 2016. It is clear that the relative humidity has the same trend as the 
temperature (see Figure \ref{fig:dailytemp}). In the summer the relative 
humidity could be as high as 65\%, while in the winter it could drop to 
35\%. The air near the ground on the Antarctic plateau is usually 
super-saturated, i.e., the ``standard'' relative humidity is larger 
than 100\% \citep{Hardy98a}. On February 3, 2015, the daily median temperature was $-38.90$\cdeg\  
and the daily median relative humidity was 60.1\%. This corresponds to ``standard'' relative humidity
 of nearly 100\%, because using Table 2 in \cite{Hardy98a}, when the ``standard'' relative humidity 
is 100\%, the relative humidity defined by WMO at $-38.90$\cdeg\ is 68\%.

\begin{figure}
\plotone{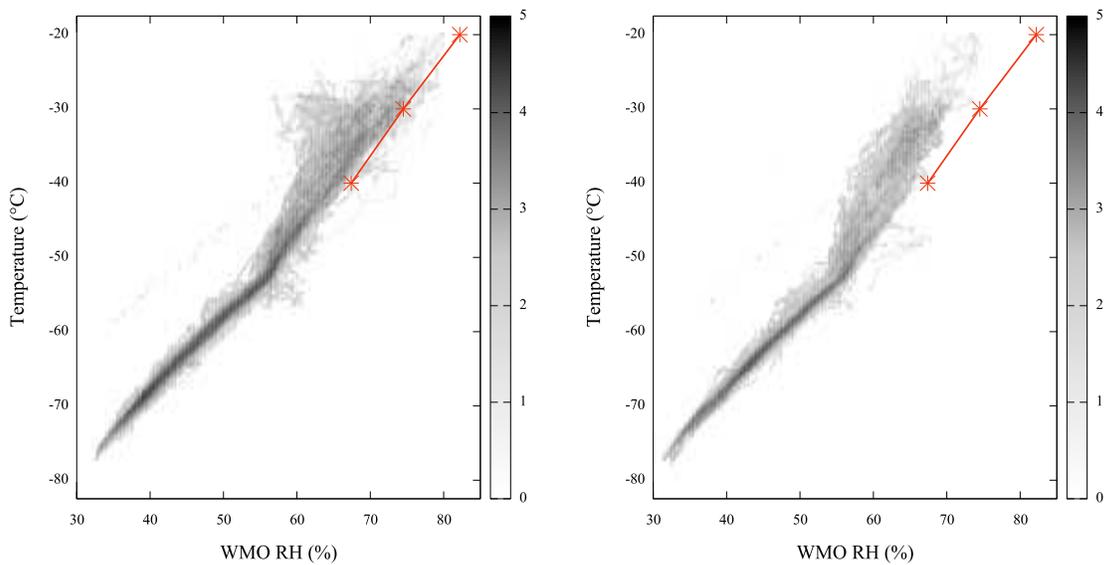}
\caption{Relationship between WMO RH and temperature during 2015 
(left) and 2016 (right). The grey scale is the logarithm of the number of 
data points in each pixel. The red line is the relationship between WMO RH and temperature 
when the ``standard" RH is 100\% given by \cite{Hardy98a}. The strong correlation indicates 
the ``standard" RH is close to 100\% all the time.
\label{fig:rhvstemp}
}
\end{figure}

Figure \ref{fig:rhvstemp} shows a strong correlation between the relative 
humidity and the temperature at 2\,m. It is obvious that there is a turnover 
near $-50$\cdeg\ in Figure \ref{fig:rhvstemp}. Based on its working 
principle, the humidity sensor measures the absolute water content in the 
air, and then translates the measurement to relative humidity by 
calculating the saturated water vapor pressure under the ambient 
temperature \citep{Hardy98b}. Our relative humidity sensor has its own 
temperature sensor, however, its measurement range is fixed from 
$-50$\cdeg\ to $+50$\cdeg. When the ambient temperature is lower than $-50$\cdeg, the 
temperature  that it reads is still $-50$\cdeg. This will overestimate 
the saturated water vapor pressure and thus result
in an underestimated relative humidity.  However, despite the turnover, the strong correlation 
shown in Figure \ref{fig:rhvstemp} still indicates a nearly 100\% ``standard'' relative humidity 
all the time at least for ambient temperature higher than $-40$\cdeg, 
and this is probably also true for lower temperature.

\section{Comparison between 2011 (KLAWS) and 2015-2016 (KLAWS-2G) data} 
\label{sec:compare}

We have collected and analyzed meteorological data from ground to 14\metre 
high at Dome~A for about three years, in 2011 with KLAWS and 2015-2016 with 
KLAWS-2G. We compare the results in order to have a better understanding of 
the site as a potential astronomical observatory.

The daily median temperatures for all the elevation above 0\metre at Dome A 
decrease rapidly from January to April as the Sun sets lower and lower, 
become quasi-constant during polar nights from May to September (although 
the day to day median temperature variation could be as large as more than 
10\cdeg), and finally rapidly go up from October to December. This trend 
is stable at Dome~A and can be clearly seen in Figure \ref{fig:dailytemp} 
during 2015 and 2016 and Figure 2 in HU14 during 2011. The same trend had 
been found at South Pole \citep{Hudson05} and Dome C \citep{Aristidi05}. 
The average temperatures of the first seven months at 2\metre and 14\metre  
are $-54$\cdeg\ and $-46$\cdeg, $-54$\cdeg\ and $-48$\cdeg, and $-55$\cdeg\ and 
$-49$\cdeg, during 2011, 2015 and 2016, respectively. This seems very 
consistent year to year.  However, the surface temperature at 0\metre 
during 2015 and 2016 could reach below $-80$\cdeg\ rarely for less than 
0.1\% of the time,  but this had never been seen in 2011 (HU14). 

The annual average wind speed at 4\metre in 2015 and 2016 are 4.2\mps\ and 
3.8\mps, respectively, much higher than that at the same elevation in 
2011, which is only 1.5\mps. Although the anemometer occasionally got 
stuck in 2011, resulting in a smaller average wind speed, this problem 
could not change the average wind speed by a factor of 3. Therefore, we 
believe that this difference reflects the annual climatological change. Another 
difference is that the wind direction exhibits a slightly preferred wind 
direction of 150\degr during 2015 and 2016, while there was no preferred 
wind direction during 2011. 

Comparing with the results we obtained in 2011 (HU14), the total fractions 
of time when temperature inversion existed above 4\metre in 2015-2016 is 
10\% to 40\% smaller than those in 2011 (see Table \ref{tab:last} and 
Table 1 in HU14). The percentages of time when temperature inversion 
duration time exceeded 25\,hr in 2015 are about 20\% smaller than those in 
2011. On the ice cap of Antarctic continent, the temperature inversion is 
driven by the difference in emissivity between the snow surface and air. It 
can be destroyed by the strong force of wind shear or net downward 
shortwave radiation flux \citep{Vignon17}. Therefore, one possible reason 
that the temperature inversion durations in 2015-2016 are shorter is the 
higher wind speeds in 2015--2016 than in 2011. However, we cannot rule out 
that the cloudy skies were more frequent during 2015--2016, which also 
generate net downward radiation flux, and thus destroy the temperature 
inversion. The large difference in temperature inversion statistics between 
2015--2016 and 2011 shows that the meteorological parameters are quite variable
from year to year. Such annual variations in temperature inversions were 
also observed by \cite{Zhou09}.

\begin{figure}
\plotone{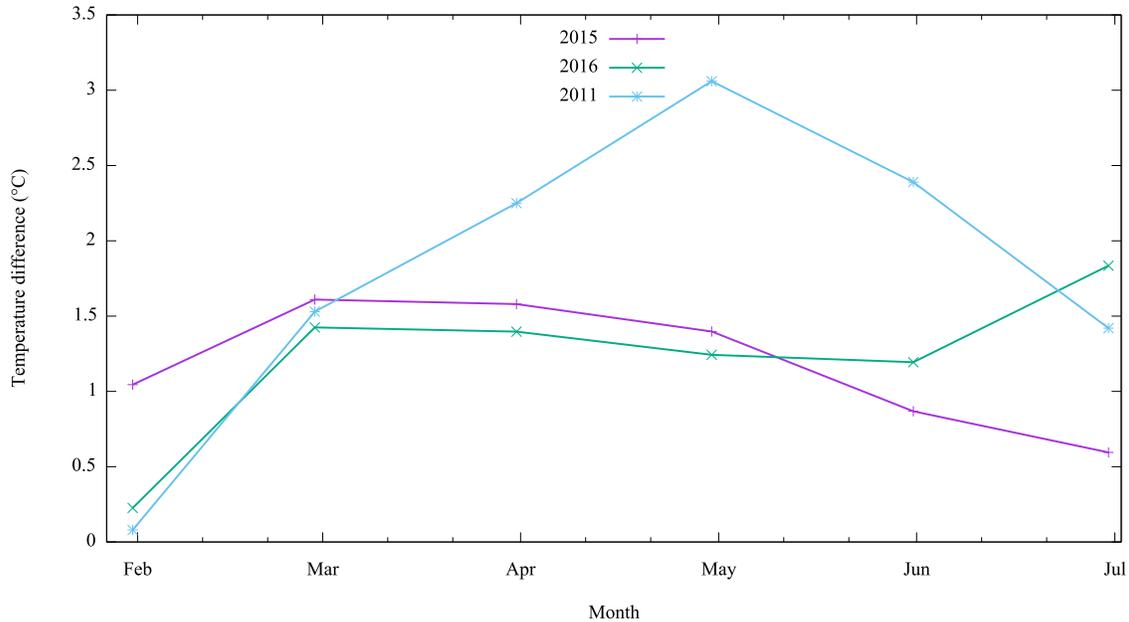}
\caption{Monthly median temperature difference at 4\metre during 2015, 2016 and 2011. 
\label{fig:monthlyvsdt}}
\end{figure}

To show the annual variation in temperature inversion, we plot  the monthly median 
temperature difference at  4\metre (e.g., $T(4m) - T(2m)$) during 2015, 2016 and  2011
in Figure \ref{fig:monthlyvsdt}. The monthly temperature difference at 4\metre during these 
three years could vary within a wide range up to 2\cdeg, especially in the winter season.

\begin{figure}
\epsscale{0.8}
\plotone{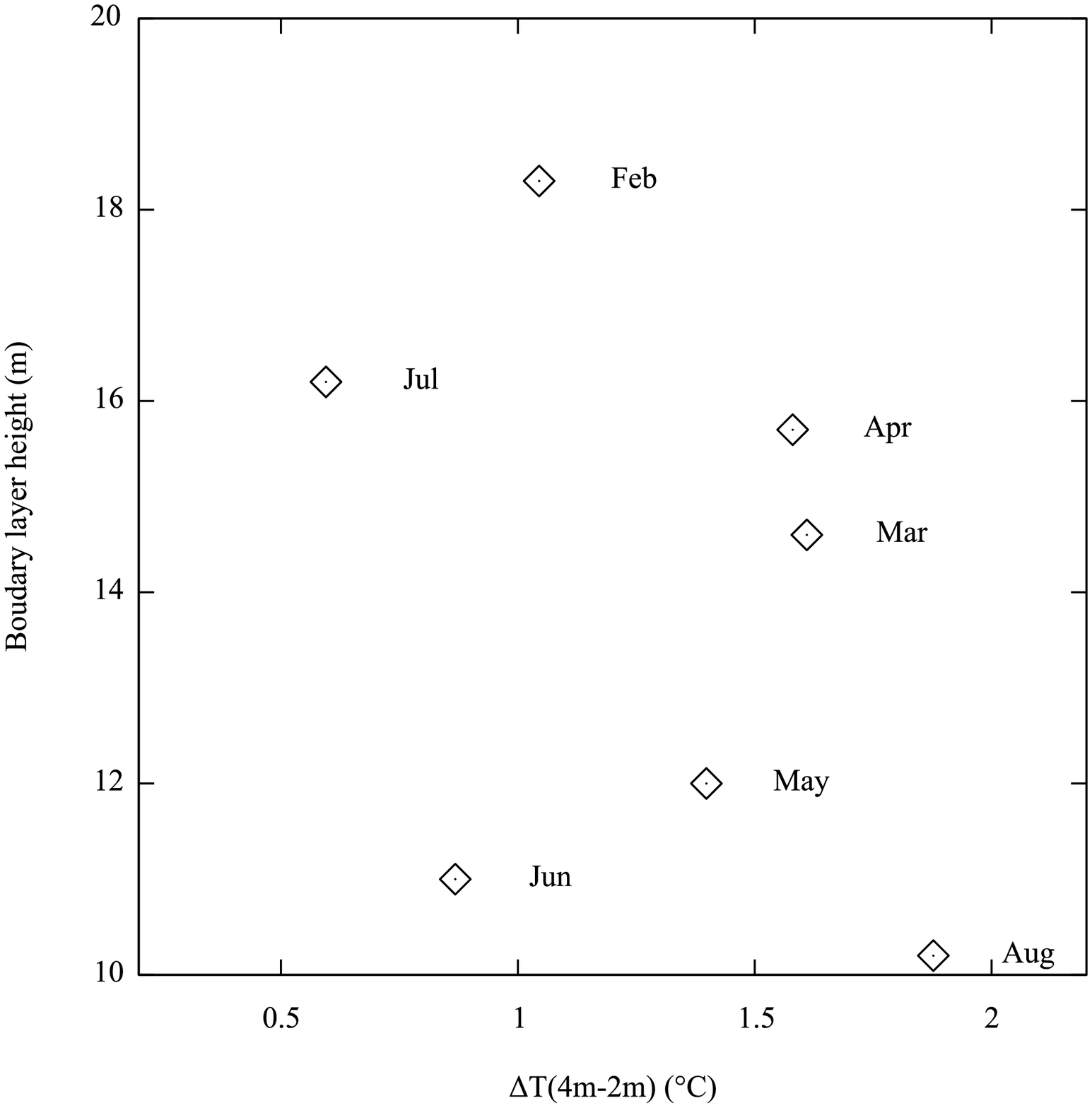}
\caption{Monthly median temperature difference at 4\metre in 2015 and 
monthly boundary layer height in 2009 \citep{Bonner10}. 
\label{fig:dtvsbl}}
\end{figure}

Similarly to HU14, we show the monthly median temperature difference at 
4\metre in 2015 and boundary layer heights in 2009 \citep{Bonner10} in 
Figure \ref{fig:dtvsbl}. The anti-correlation is not as obvious as we had 
found in 2011 (Figure 22 in HU14). This could attribute to the yearly 
variation (see Figure \ref{fig:monthlyvsdt}).

\begin{figure}
\plotone{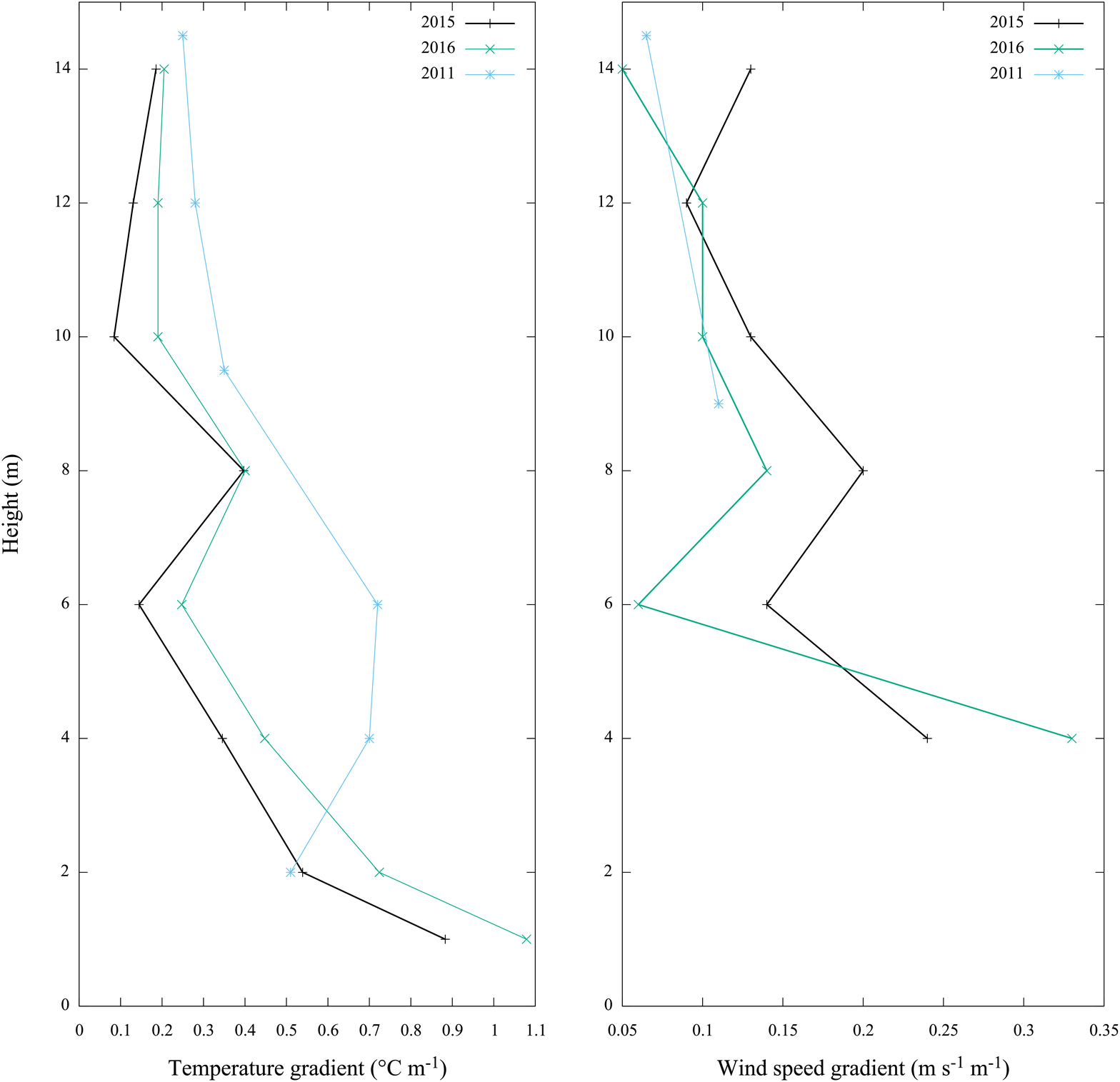}
\epsscale{0.8}
\caption{Left panel: temperature gradient profile. Right panel: wind speed 
gradient profile. The temperature gradients are calculated by corrected temperature at 10\metre
after 1 October 2015 (see Section \ref{sec:temp}). \label{fig:profile}}
\end{figure}

Finally, we plot median temperature gradient profile and median wind speed 
gradient profile in Figure \ref{fig:profile}. We note that the temperature 
gradients at 10\metre and 12\metre was obtained using corrected 10\metre temperature 
(see Section \ref{sec:temp}). Temperature gradient profiles in 2015 and 2016 
exhibit similar behavior at the lower elevations, while the profile in 2011 
is completely different. The temperature gradients decrease dramatically 
from 1\metre to 6\metre in 2015 and 2016, and then increase moderately at 
8\,m. The wind speed gradient profiles also have such turnover at 6\metre 
(see right panel in Figure \ref{fig:profile}). However, both temperature 
gradient and wind speed gradient above 8\metre asymptotically approach 
positive constant values, at least in 2011 and 2015. 

\section{Summary and Discussion}\label{sec:summary}

By analyzing data collected from KLAWS-2G at Dome A during 2015 and 2016, 
we find that the temperature seems to have a similar seasonal trend from 
year to year, which is reasonable. The average wind speed at 4\metre in 
2015--2016 was around 4\mps, much higher than that (1.5\mps) in 2011. 
Nevertheless, these are still very low wind speeds compared with that at 
temperate astronomical observatories. Unlike in 2011, there was a slightly 
preferable wind direction of 150\degr in 2015--2016. All the differences 
indicate the existence of annual variations in the meteorological 
parameters at Dome~A.

Most importantly, we have still found there is a very strong temperature 
inversion (temperature gradient reaching up to 7\cgrad at 4\,m)  at all the 
heights above the snow surface, confirming our results in 2011 (HU14). Our 
calibration results of the temperature sensors in Section 
\ref{subsec:calib} have proved that such a temperature inversion, as we 
defined as positive temperature difference larger than 0.14\cdeg, is a 
real phenomenon rather than inconsistency between temperature sensors. 
Whenever a temperature inversion exists, competition between shear and 
buoyancy will weaken turbulence and result in a shallower boundary layer. 
Thus the atmospheric seeing will probably become better when temperature 
inversion occurs.  We have also built a similar AWS and installed it at 
Muztagata site in Pamirs Plateau in April 2017. DIMMs have been installed 
at that site for directly measuring atmospheric seeing. Although 
temperature inversion is not as frequent or strong as that at Dome~A, when 
it exists, the seeing is usually better (private communication).

\begin{figure}
\epsscale{0.8}
\plotone{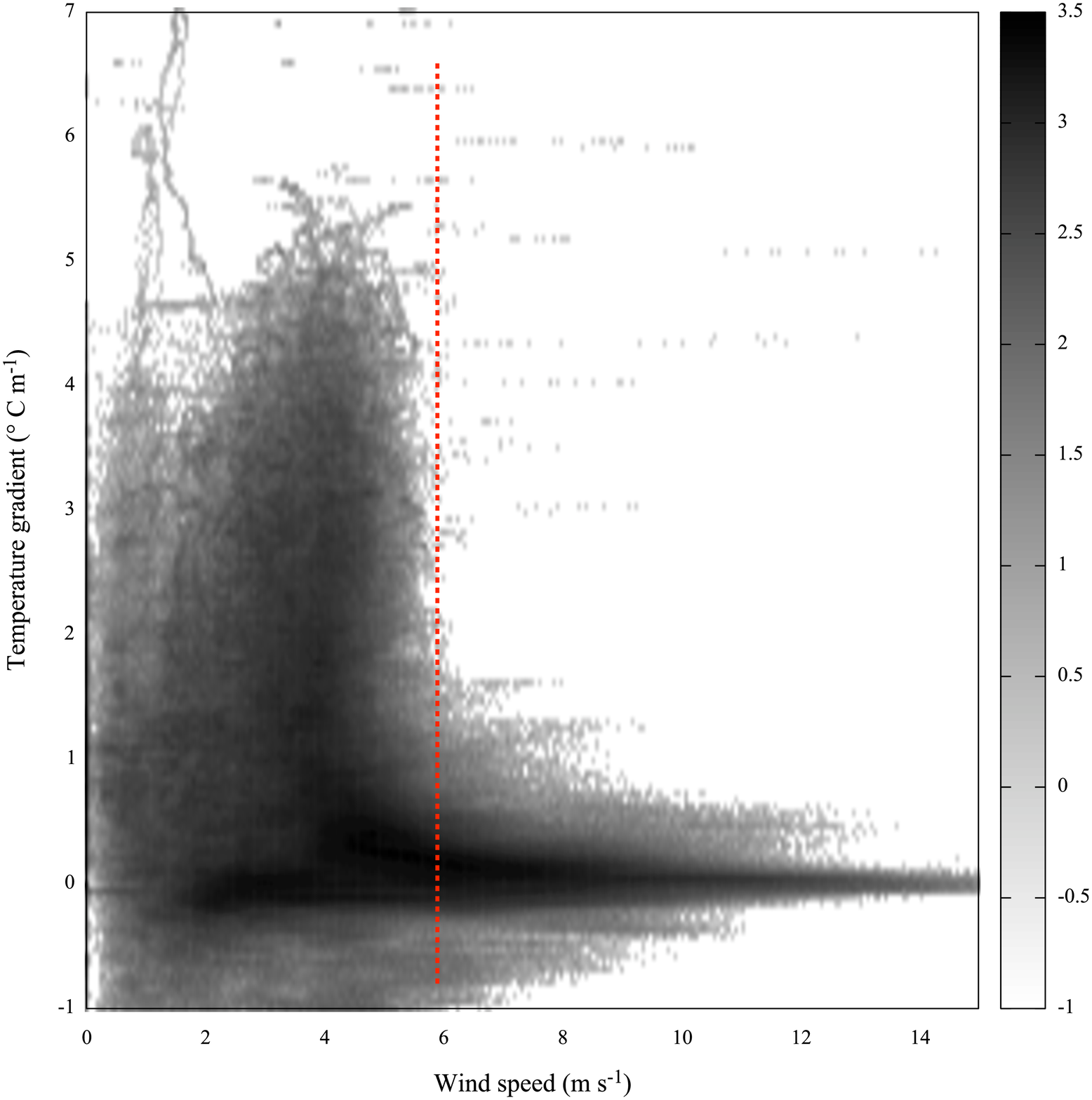}
\caption{Wind speed at 4\metre versus temperature gradient at 4~m. The grey 
scale is the logarithm of the number of data points per pixel. The vertical 
red dash line at 6.0\mps\ indicates the threshold wind speed. 
\label{fig:wsvsgradient}}
\end{figure}

By studying meteorological data at Dome C, \cite{Vignon17} claimed that the 
stable boundary layer (SBL) could be divided into two regimes with respect 
to the wind speed at 10\,m. In the first regime, the wind speed is strong, 
associated with continuous turbulence. In the second regime, the wind speed 
is weak, the temperature inversion is strong and thus the turbulent 
activity is weak. Figure \ref{fig:wsvsgradient} shows the relation between 
wind speed and temperature gradient at 4\metre in 2015. It is clear that 
there is a wind speed threshold roughly around 6.0\mps. Below the 
threshold, the temperature gradients are scattered over a very wide range 
and could reach as large as 7\cgradd. When the wind speeds are stronger 
than 6.0\mps, the temperature gradients are moderate and their absolute 
values slightly decrease with the wind speed. The relation between wind 
speed and temperature gradient in Figure \ref{fig:wsvsgradient} implies 
that there are also two distinct SBL regimes at Dome A which could be 
characterized by the local wind speed.  

Large atmospheric seeing within a narrow turbulent boundary layer is ubiquitous on Antarctica 
continent, as observed, e.g., by \cite{Marks96} and \cite{Aristidi09}, and theoretically studied by \cite{Swain06}. Therefore, 
a telescope should be sited above the boundary layer to avoid the strong turbulence in the ground layer. 
\cite{Vignon17} found the temperature profile near the ground was a 
convex-concave-convex shape in the first SBL regime at Dome C and an 
exponential shape in the second SBL regime. However, the median temperature 
gradient profile at Dome~A (left panel in Figure \ref{fig:profile}) is 
somehow different from an exponential shape, since the 
convex-concave-convex shaped profile in the first regime also contributes 
to the median gradient profile. There might be a critical elevation at Dome 
A, above which the temperature gradient is asymptotic to positive constant. 
The occurrence of small positive constant temperature gradient indicates 
the air in that elevation tends to be in equilibrium state. This critical 
elevation might be the optimized height to avoid the relatively strong 
turbulence near the surface. As we mentioned above, the temperature 
gradient profile below 8\metre varies significantly year by year, but the 
temperature gradient asymptotically approaches a constant above 8\,m. 
Therefore, 8\,m is a minimum height for building a telescope at Dome A. 
We will verify this with direct atmospheric seeing measurements 
by DIMM.

In conclusion, we find that strong temperature inversions (temperature 
gradient reaching up to 7\cgrad at 4\,m) existed for all the heights 
above the ground at Dome A during 2015 and 2016. The temperature inversion, 
along with a moderate wind speed (averagely 4.2\mps), produces a shallow 
boundary layer. It would imply that there will be a considerable percentage of 
time when a telescope that is placed just 8 to 10\metre above the ground 
can obtain superior free-atmospheric seeing.

However, the complex temperature and wind speed gradient profiles imply 
that small-scale simulation should be performed for studying the turbulence 
below the boundary layer. Also, comparing with data from KLAWS in 2011, we 
find annual variations in temperature, temperature inversion and wind 
speed at Dome A. Therefore, long-term and continuous data are needed for 
thoroughly understanding the site characteristics at Dome A. We plan to
install a new identical KLAWS-2G for long-term monitoring at Dome A in early 2019.

The data from KLAWS-2G during 2015 and 2016 are available at 
\url{http://aag.bao.ac.cn/klaws/downloads} and as plots at 
\url{http://aag.bao.ac.cn/klaws}.

\acknowledgments

The authors wish to thank all members of the 31st, 32nd and 33rd Chinese 
Antarctic Research Expedition teams for their effort in setting up the 
KLAWS-2G instruments and servicing the PLATO-A observatory. This research is 
supported by the National Natural Science Foundation of China under grant 
Nos. 11403048, 11733007, and 11673037, the Chinese Polar Environment 
Comprehensive Investigation \& Assessment Programmes under grant No. 
CHINARE2016-02-03, the National Basic Research Program of China (973 
Program 2013CB834900), the Australian Antarctic Division, and the Australian 
National Collaborative Research Infrastructure Strategy administered by 
Astronomy Australia Limited.

\end{document}